\documentclass[twocolumn]{aastex62}
\usepackage{threeparttable}
\graphicspath{{./}{figures/}}

\received{November 2, 2020}
\revised{February 1, 2021}
\accepted{February 16, 2021}

\shorttitle{NIR Spectroscopy of 55 Cancri e}
\shortauthors{Deibert et al.}

\usepackage{aas_macros}
\usepackage{natbib}
\begin{document}

\title{A Near-Infrared Chemical Inventory of the Atmosphere of 55 Cancri e}

\correspondingauthor{Emily K. Deibert}
\email{deibert@astro.utoronto.ca}

\author[0000-0001-9796-2158]{Emily K. Deibert}
\affil{David A. Dunlap Department of Astronomy \& Astrophysics, University of Toronto, 50 St. George Street, ON M5S 3H4, Canada}
\affil{Dunlap Institute for Astronomy \& Astrophysics, University of Toronto, 50 St. George Street, ON M5S 3H4, Canada}

\author{Ernst J. W. de Mooij}
\affil{Astrophysics Research Centre, Queen's University Belfast, Belfast BT7 1NN, UK}

\author[0000-0001-5349-6853]{Ray Jayawardhana}
\affil{Department of Astronomy, Cornell University, Ithaca, New York 14853, USA}

\author{Andrew Ridden-Harper}
\affil{Department of Astronomy, Cornell University, Ithaca, New York 14853, USA}

\author{Suresh Sivanandam}
\affil{David A. Dunlap Department of Astronomy \& Astrophysics, University of Toronto, 50 St. George Street, ON M5S 3H4, Canada}
\affil{Dunlap Institute for Astronomy \& Astrophysics, University of Toronto, 50 St. George Street, ON M5S 3H4, Canada}

\author{Raine Karjalainen}
\affil{Astronomical Institute, Czech Academy of Sciences, Fri\v{c}ova 298, 25165, Ond\v{r}ejov, Czech Republic}
\affil{Instituto de Astrof\'{i}sica de Canarias, c/ V\'{i}a L\'{a}ctea s/n E-38205 La Laguna, Tenerife, Spain}
\affil{Isaac Newton Group of Telescopes, Apartado de Correos 321, Santa Cruz de La Palma, E-38700, Spain}

\author{Marie Karjalainen}
\affil{Astronomical Institute, Czech Academy of Sciences, Fri\v{c}ova 298, 25165, Ond\v{r}ejov, Czech Republic}

\begin{abstract}
We present high-resolution near-infrared spectra taken during eight transits of 55~Cancri~e, a nearby low-density super-Earth with a short orbital period ($<$ 18 hours). While this exoplanet's bulk density indicates a possible atmosphere, one has not been detected definitively. Our analysis relies on the Doppler cross-correlation technique, which takes advantage of the high spectral resolution and broad wavelength coverage of our data, to search for the thousands of absorption features from hydrogen-, carbon-, and nitrogen-rich molecular species in the planetary atmosphere. Although we are unable to detect an atmosphere around 55 Cancri e, we do place strong constraints on the levels of HCN, NH${}_3$, and C${}_2$H${}_2$ that may be present. In particular, at a mean molecular weight of 5 amu we can rule out the presence of HCN in the atmosphere down to a volume mixing ratio (VMR) of 0.02\%, NH${}_3$ down to a VMR of 0.08\%, and C${}_2$H${}_2$ down to a VMR of 1.0\%. If the mean molecular weight is relaxed to 2 amu, we can rule out the presence of HCN, NH${}_3$, and C${}_2$H${}_2$ down to VMRs of 0.001\%, 0.0025\%, and 0.08\% respectively. Our results reduce the parameter space of possible atmospheres consistent with the analysis of HST/WFC3 observations by \cite{Tsiaras16}, and indicate that if 55~Cancri~e harbors an atmosphere, it must have a high mean molecular weight and/or clouds.
\end{abstract}



\section{Introduction} \label{sec:intro}
Over the past several years, our understanding of hot Jupiter atmospheres has expanded enormously. This is due in large part to high-precision observations of transiting exoplanets that have enabled the detection of atomic and molecular species and provided constraints on the atmospheric structures of several gas giants (see e.g. \citealt{Madhusudhan19} for a broad overview of previous detections). 

In contrast, the atmospheric properties of lower-mass, super-Earth exoplanets remain largely unconstrained. Their shallower transit depths and smaller atmospheric scale heights produce weaker spectroscopic signals that are more challenging to detect given our current observational capabilities. However, these atmospheres are of great scientific interest: in particular, they are predicted to be extraordinarily diverse, potentially being rich in carbon, silicate, and/or water vapors \citep[e.g.][]{Schaefer09,Miguel11,Hu14}. Their  atmospheric compositions are likely to reflect the varied formation and evolutionary histories that exoplanets in the super-Earth regime have experienced \citep[e.g.][]{Madhusudhan16}, and can shed light on these otherwise-elusive worlds.

A super-Earth of particular interest is the nearby transiting planet 55 Cancri e (hereafter referred to as 55~Cnc~e), the existence of which was first suggested by \cite{McArthur04}. \cite{Dawson10} later determined that its initially-derived period of 2.808 days was an alias of its true, much shorter period of $\sim$ 18 hours; this value was recently refined by \cite{Bourrier18}. 55~Cnc~e has a mass of $\sim 8 M_\oplus$ and radius of $\sim 2 R_\oplus$ (\citealt{Crida18}). It orbits a bright (V=5.95; \citealt{vonBraun11}) G8V host star.
The ultra-short orbital period of 55~Cnc~e results in an equilibrium temperature in excess of 2000K \citep[e.g.][]{Demory16}, potentially leading to exotic atmospheric properties.

While the planet's bulk density indicates that it could harbor an atmosphere (see for e.g. \citealt{Gillon12, Bourrier18}), several observational attempts have not been able to definitely detect its presence. In particular, \cite{Ehrenreich12} found no evidence for an extended hydrogen atmosphere, and \cite{Esteves17} and \cite{Jindal20} derived limits on water absorption consistent with the exoplanet having either a hydrogen-poor atmosphere or a hydrogen-rich atmosphere that is significantly depleted in water vapour. 
\cite{Demory16} measured the photometric phase curve of 55~Cnc~e at 4.5 $\mu$m with the Spitzer Space Telescope, finding a large temperature contrast between the exoplanet's permanent day and night sides and a hot-spot on the day side offset by 40$^\circ$ from the substellar point. 

Using grism spectroscopy data from the Wide Field Camera 3 (WFC3) aboard the Hubble Space Telescope (HST), \cite{Tsiaras16} reported the detection of an atmosphere around 55~Cnc~e and suggested that it is likely hydrogen-rich, with a large scale height and high C/O ratio. They indicate that HCN is the most likely molecular candidate able to explain features detected at 1.42 and 1.54 $\mu$m, but caution that additional observations over a broader wavelength range would help confirm the results.

\cite{Hammond17} modelled the phase curve of 55~Cnc~e using an atmospheric global circulation model (GCM) and found that a 90\% -- 10\% mixture of H$_2$ and N$_2$ in the atmosphere with cloud-forming species such as SiO can well approximate the observed phase variations. However, a complementary analysis by \cite{Angelo17} found that the atmosphere is likely dominated by either CO or N$_2$ with minor abundances of H$_2$O or CO$_2$. More recently, \cite{Miguel19} explored the expected chemical composition of the atmosphere of 55~Cnc~e, and concluded that transmission spectra should show strong features of NH${}_3$ and HCN at mid- to long-infrared wavelengths if the atmosphere is nitrogen-rich, as may be expected from the large day-night temperature contrast \citep{Hammond17}.

In this paper, we present high-resolution spectroscopy of 55~Cnc~e from 950 -- 2350 nm, focusing in particular on absorption features in the near-infrared (NIR) due to HCN.
We also investigate the presence of NH${}_3$, C${}_2$H${}_2$, CO, CO${}_2$, and H${}_2$O at NIR wavelengths. Our search for HCN is informed by the observational results of \cite{Tsiaras16}, who suggested that HCN is present in the atmosphere at a high volume mixing ratio (VMR) with acceptable values as low as $10^{-5}$.
In addition, \cite{Zilinskas20} explored plausible VMRs for the case of a nitrogen-dominated atmosphere \cite[following][]{Miguel19}. They find that CO may be present at relatively high VMRs ($\sim10^{-4.3}$, their Fig. 3) across a wide range of pressures and C/O ratios. The expected VMR of H${}_2$O depends largely on the C/O ratio, while CO${}_2$ is expected to have a VMR $\lesssim10^{-7.5}$ for all pressures and C/O ratios considered. C${}_2$H${}_2$ is only expected at VMRs $>10^{-8}$ for C/O $>$ 2; however, tentative results from \cite{Tsiaras16} suggested that C${}_2$H${}_2$ may be present with a VMR as high as $10^{-5}$ at C/O $\sim$ 1.1. \cite{Zilinskas20} also note that if nitrogen is introduced to the system, the atmosphere will tend to form HCN even if hydrogen is only present in small amounts.

Our analysis takes advantage of two facets of our observations: first, the wide wavelength coverage of our data,
which span thousands of absorption features of both water and various carbon- and nitrogen-rich molecules in the NIR; and second, the very high spectral resolving powers (R $\sim$ 80,000) of our observations, which allow us to individually resolve these thousands of absorption features. Combining these features allows us to increase our detection capabilities, shedding further light on the atmospheric composition of 55~Cnc~e.

Our paper is structured as follows. In Section \ref{sec:obs}, we describe the eight nights of observations obtained for this exoplanet, and in Section \ref{sec:reduc}, we summarize the data reduction process for these observations. Our analysis and results are presented in Sections \ref{sec:analysis} and \ref{sec:discussion}, and we discuss these results in Section \ref{sec:actual_discussion}. Our conclusions follow in Section \ref{sec:conclusion}, with additional details in the appendices.

\section{Observations} \label{sec:obs}
We observed six transits of 55~Cnc~e with the Calar Alto high-Resolution search for M dwarfs with Exoearths with Near-infrared and optical \'{E}chelle Spectrographs (CARMENES; \citealt{carmenes}) located at the Calar Alto Astronomical Observatory. In this paper we focus on the NIR channel, which spans the wavelength range of 960 -- 1710 nm.

\begin{deluxetable*}{cccccccc}
\tabletypesize{\footnotesize}
\tablecaption{Summary of observations.
\label{tab:obs}}
\tablehead{%
    \colhead{Night} &\colhead{Date (UT)} & \colhead{Instrument/Telescope} & \colhead{Duration (hr)} & \colhead{Frames (In/Out)\tablenotemark{1}} & \colhead{Exp. Time (s)\tablenotemark{2}} & \colhead{Avg. SNR} & \colhead{Used in Analysis?}
    }
\startdata
1 & 2016 Dec 26 & CARMENES/Calar Alto & 11.6 & 363 (80/283) & 33.9 & 91 & Y \\
2 & 2017 Nov 22 & CARMENES/Calar Alto & 5.7 & 141 (55/86) & 58.0 & 145 & Y \\
3 & 2017 Nov 24 & CARMENES/Calar Alto & 5.8 & 115 (43/72) & 87.8 & 157 & Y \\
4 & 2017 Dec 9 & CARMENES/Calar Alto & 1.0 & 26 (9/16) & 53.1 & 108 & N \\
5 & 2017 Dec 11 & CARMENES/Calar Alto & 5.6 & 92 (45/47) & 86.6 & 71 & Y \\
6 & 2017 Dec 17 & CARMENES/Calar Alto & 5.5 & 59 (14/45) & 87.9 & 74 & N \\
7 & 2019 Feb 14 & SPIRou/CFHT & 2.6 & 77 (47/30) & 94.7 & 268 & Y \\
8 & 2019 Feb 25 & SPIRou/CFHT & 2.5 & 73 (47/26) & 94.7 & 300 & Y \\
9 & 2019 Apr 17 & SPIRou/CFHT & 2.7 & 54 (34/20) & 94.7 & 247 & Y \\
10 & 2019 May 1 & SPIRou/CFHT & 2.3 & 73 (46/27) & 94.7 & 250 & Y \\
\enddata
\tablenotetext{1}{(In/Out) refers to the number of in- and out-of-transit frames for each night.}
\tablenotetext{2}{The exposure times refer to the average exposure time across all observations for each night.}
\end{deluxetable*}

The observations were taken with varying exposure times (see Table \ref{tab:obs}), and in general each observation covered one full transit of 55~Cnc~e (the transit duration being approximately 1.6 hours). We note that the observations taken on Night 4 and Night 6 (hereafter N${}_4$ and N${}_6$, respectively) suffered from poor observing conditions and did not cover the full transit. For this reason, we chose to exclude these nights from our analysis. The spectral resolving power of the instrument in the NIR is nominally $\lesssim$ 80,000 (\citealt{Quirrenbach18}).

We also observed four transits of 55~Cnc~e with the SpectroPolarim\`{e}tre Infra-Rouge (SPIRou; \citealt{spirou}) located at the Canada France Hawai'i Telescope (CFHT). The wavelength coverage of the data is 950 -- 2350 nm.
The 2-3h observation period of each night covered one full transit of 55~Cnc~e, for a total of four full transits. The spectral resolving power of the instrument is nominally $\sim$~75,000 (\citealt{spirou}).

In total, we observed 8 full transits and 2 partial transits of 55~Cnc~e. The 2 partial transits suffered from poor observing conditions and were excluded from further analysis; thus, this paper reports on our analysis of 8 full transits.
Each night also includes a baseline of out-of-transit data, which is used to assess the significance of our results (see Section \ref{subsec:significance}).
A summary of our observations can be found in Table \ref{tab:obs}, and a list of the parameters used in this analysis can be found in Table \ref{tab:parameters}.

\begin{deluxetable}{lcc}
\tablecaption{Stellar and Planetary Parameters Used in This Analysis}
\tablehead{%
    \colhead{Parameter} & \colhead{Value} & \colhead{Reference}
    }
\startdata
Spectral Type & G8 & \cite{vonBraun11} \\
J band Flux & 4.59 & \cite{2002yCat.2237....0D} \\
H band Flux & 4.14 & \cite{2002yCat.2237....0D} \\
K band Flux & 4.015 & \cite{2003yCat.2246....0C} \\
$T_\mathrm{eff}$ (K) & 5172 & \cite{Yee17} \\
log $g$ & 4.43 & \cite{Yee17} \\
$[$Fe/H$]$ & 0.35 & \cite{Yee17} \\
$R_* \text{ } (R_\odot)$ & 0.980 & \cite{Crida18} \\
$K_*\tablenotemark{1} \text{ } (\text{m } \text{s}^{-1})$ & 6.02 & \cite{Bourrier18} \\
$m_p \text{ } (M_\oplus)$ &  8.59 & \cite{Crida18} \\
$r_p \text{ }(R_\oplus)$ & 1.947 & \cite{Crida18} \\
$V_{\text{sys }} (\text{km } \text{s}^{-1}$) & 27.58 & \cite{Nidever2002} \\
Orbital Period (days) & 0.7365474 & \cite{Bourrier18} \\
Mid-transit (JD) & 2457063.2096 & \cite{Bourrier18} \\
Semimajor Axis (au) & 0.01544 & \cite{Bourrier18} \\
Inclination (degrees) & 83.59 & \cite{Bourrier18} \\
$R_p/R_*$ & 0.0182 & \cite{Bourrier18} \\
$a/R_*$ & 3.52 & \cite{Bourrier18} \\
$\mu_1$ & 0.3156 & \cite{Claret2004} \\
$\mu_2$ & 0.2893 & \cite{Claret2004} \\
\enddata
\tablenotetext{1}{$K_*$: radial velocity semi-amplitude}
\label{tab:parameters}
\end{deluxetable}

\section{Data Reduction} \label{sec:reduc}

\begin{figure*}
    \centering
    \includegraphics{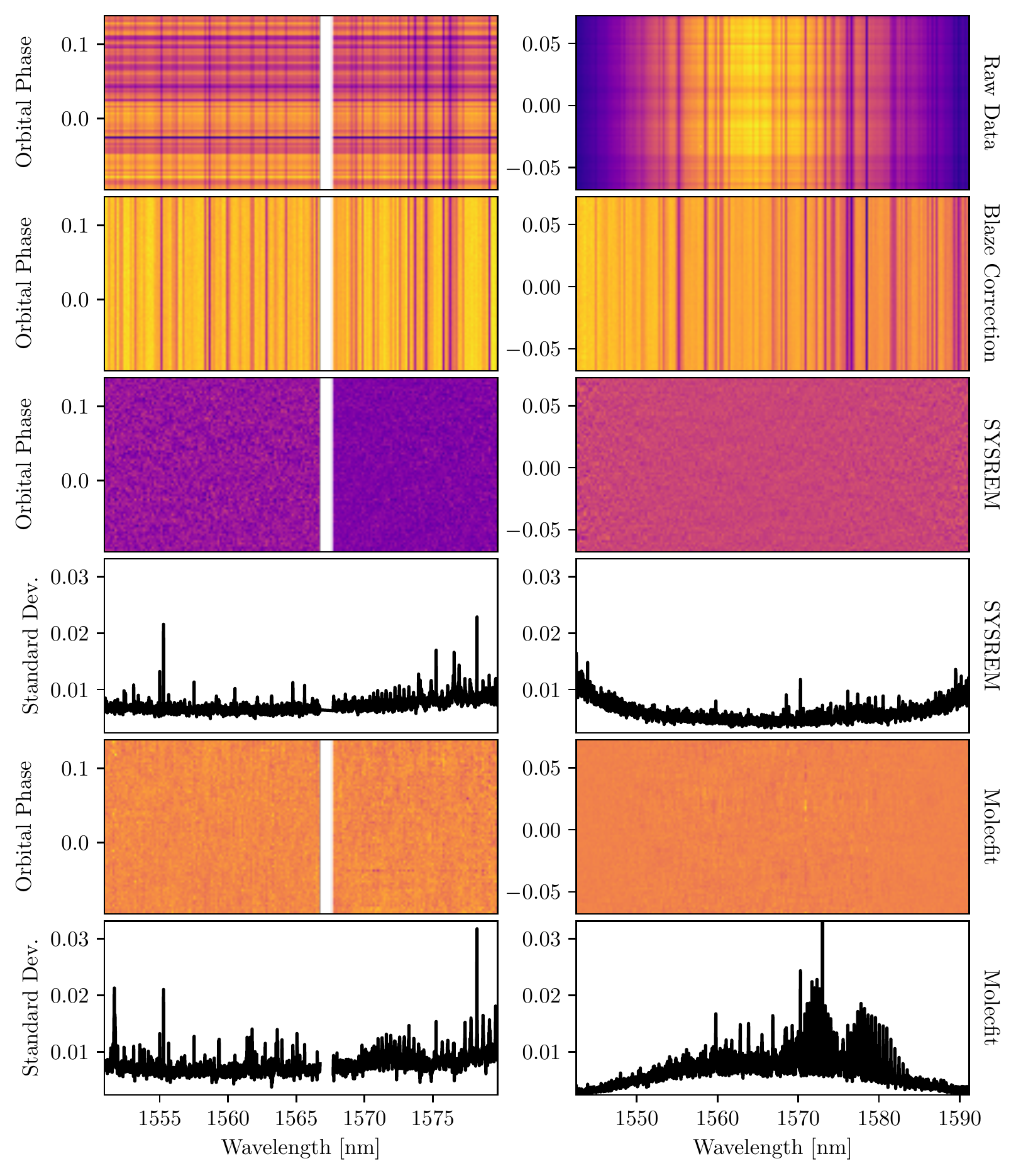}
    \caption{An example of the data reduction process described in Section \ref{sec:reduc}, as applied to a single order of Night 2 of the CARMENES data (left) and the SPIRou data (right). In each case, the top panel shows the raw data; the second panel shows the data after the blaze correction and removal of cosmic rays as described in Section \ref{sec:reduc}; the third panel shows the data after 6 iterations of the \textsc{sysrem} algorithm (see Section \ref{subsec:sysrem}); the fourth panel shows the standard deviation across each wavelength channel after applying the \textsc{sysrem} algorithm; the fifth panel shows the data (i.e. the second panel) after correcting it with Molecfit (see Section \ref{subsubsec:molec}); and the sixth panel shows the standard deviation along each wavelength channel after correcting with Molecfit and removing residual stellar lines. Note that there are gaps present in the CARMENES orders; this is visible in the left panels.}
    \label{fig:reduction}
\end{figure*}

The raw data frames obtained using CARMENES were initially reduced by the observatory with the CARMENES reduction pipeline CARACAL (\citealt{Caballero16}), while the data obtained using SPIRou were extracted by the observatory using the SPIRou Data Reduction Software (DRS)\footnote{ \url{http://www.cfht.hawaii.edu/Instruments/SPIRou/SPIRou\_pipeline.php}}. Note that while the SPIRou DRS does include a telluric correction, we have chosen to carry out this correction ourselves (see Section \ref{subsec:sysrem}); the data products extracted from the SPIRou DRS are thus wavelength-corrected 2D spectra. Note as well that both pipelines supply a Barycentric Earth Radial Velocity (BERV) correction. Examples of extracted spectra for both CARMENES and SPIRou are shown in the top left and top right panels of Fig. \ref{fig:reduction} respectively, while all extracted and reduced spectra are shown in the first panels of the figures presented in Appendix \ref{app:reduc}.

The next steps of our data reduction process proceeded similarly to \cite{Deibert19}. We interpolated our data to a common wavelength grid with a linear interpolation. We then removed contaminating comic rays through median filtering, where points falling outside of a threshold of 5 median absolute deviations were flagged as outliers and masked during subsequent analyses.

To account for the grating-dependent variation in brightness at different spectral orders in the data (i.e. the ``blaze function''), we carried out a blaze correction. Such a correction is necessary for observations taken using \'{E}chelle spectrographs. The effect was removed separately for each night and each individual order of the data. To do this, we divided the first science spectrum of each night (the ``reference spectrum'') from all science spectra, and fit this product with a low-order polynomial. The polynomial fit was then divided out of each individual observation.
This removed the effects of the blaze response function, resulting in a normalized spectrum that could be used for subsequent analysis. The procedure additionally corrects for other wavelength-dependent variations such as airmass effects or differential slit losses.

The results of these initial data reduction steps on a single order of N${}_2$ (the 24th order) and N${}_8$ (the 30th order) can be seen in the second left and second right panels of Fig. \ref{fig:reduction} for the CARMENES and SPIRou observations, respectively. Additionally, the results of applying these corrections to all nights/orders of both datasets are shown in the second panels of the figures presented in Appendix \ref{app:reduc}.

\subsection{Correction of Systematic Effects}
\label{subsec:sysrem}
After the initial reduction process described above, the resultant spectra were largely dominated by both stellar and telluric absorption lines, seen as vertical lines in, e.g., the second left and second right panels of Fig. \ref{fig:reduction}. However, due to the essentially-stationary nature of these lines when compared with absorption due to the exoplanet's atmosphere (which varies in radial velocity from $\sim -60$ km/s to $\sim +60$ km/s over the course of the transit) these features can be removed using the \textsc{sysrem} algorithm, first described by \cite{Tamuz2005}. \textsc{sysrem} allows for linear, systematic effects that appear in many datasets of the same sample to be removed through an algorithm that, in the case of equal uncertainties, reduces to a Principal Component Analysis (PCA) algorithm. 

The correction was determined using both in- and out-of-transit frames, and each order of the data was treated separately. For all observations, the average airmass at each exposure was taken as an initial approximation of the systematic effects to be removed. For consistency, we chose to apply the same number of iterations of the algorithm to each order; to determine this number, we calculated the root-mean-square (rms) of the residuals after subsequent iterations of \textsc{sysrem}. We found that on average, six iterations of the algorithm was the point at which the rms of the residuals had plateaued and additional applications of the algorithm did not yield a significant improvement. The third left and third right panels of Fig. \ref{fig:reduction} show the results of applying 6 iterations of the \textsc{sysrem} algorithm to CARMENES and SPIRou spectra, respectively. The results of applying six iterations to all nights/orders of each data set are shown in the figures in Appendix \ref{app:reduc}.

The algorithm performs poorly around strong or closely-spaced lines; in particular, several orders of our observations (i.e. those at the edges of the Y, J, H, and K photometric bands) suffer from significant telluric absorption. These strong lines make the blaze response correction difficult, and as a result, nonlinear effects are introduced into the spectra which cannot be removed using \textsc{sysrem}. To reduce contamination from these poorly-constrained frames, we followed e.g. \cite{Snellen10}, \cite{Esteves17}, and \cite{Deibert19} in weighting each frame and each pixel by its standard deviation (shown in the fourth panels of Figs. \ref{fig:reduction}, as well as the figures presented in Appendix \ref{app:reduc}). This step reduces the contribution from the noisier portions of the data. Additionally, we have chosen to fully exclude several particularly-contaminated orders (those located within the gaps between the Y, J, H, and K bands) from further analysis. These are described in greater detail in Appendix \ref{app:reduc}.

\subsubsection{Telluric Feature Removal with Molecfit}
\label{subsubsec:molec}

Although the use of \textsc{sysrem} for the removal of telluric features is well-established (see for e.g. the discussion in \citealt{Birkby18}), we repeated the telluric removal process for a subset of our observations/atmospheric models using Molecfit \citep{Smette15,Kausch15}, a tool originally developed for the European Southern Observatory (ESO) consortium that corrects telluric absorption based on synthetic modelling of the Earth's atmospheric transmission.
Molecfit has recently been used on both high-resolution optical and NIR observations (e.g., \citealt{Allart17,Allart18,Salz18,Cabot20,Seidel20}, among others). 

Here, we largely followed the methods laid out in \cite{Allart17} and \cite{Salz18} to apply Molecfit to our own observations. We used version 2.0.1 of the software, which was the most recent version available at the time of our analysis. Following \cite{Salz18}, we used a high-resolution synthetic stellar spectrum (PHOENIX; \citealt{Husser13}) matching the properties of 55 Cnc A ($T_\mathrm{eff} \approx 5200$K, log $g \approx$ 4.5 dex, $\sim$ solar metallicity; \citealt{Yee17}) in order to {identify and} mask stellar features in our spectra. {We then checked these masked regions by eye, and adjusted where necessary.} {Following this, we} selected narrow regions containing intermediate-strength telluric lines (i.e. not saturated), {few or no stellar lines,} and a flat continuum distributed throughout our spectra to fit. The remaining regions of each spectrum were corrected with the Calctrans tool, which is used to apply fits from Molecfit to the rest of the data. Each individual spectrum was fit and corrected separately. The goodness-of-fits for our data are comparable to those obtained by \cite{Allart17}; in our case, the average $\chi^2$ across all fits is 10.9. We note that this somewhat large value can likely be explained by the fact that our fit does not include stellar lines. Examples of this correction can be seen in the fifth panels of Fig. \ref{fig:reduction}, while an example of the standard deviation of this correction across wavelength channels can be seen in the final panels of Fig. \ref{fig:reduction}.

Additional details on our fit results, as well as the parameters used for our fits, are presented in Appendix \ref{subapp:params}. We discuss the efficacy of Molecfit compared to that of \textsc{sysrem} in further detail in Section \ref{subsec:comparison}.

\section{Analysis} \label{sec:analysis}

Our analysis relies on the Doppler cross-correlation technique, which was used to obtain a robust atmospheric detection by \cite{Snellen10}. An overview of the work preceding \cite{Snellen10}, as well as the numerous detections made since then, is found in \cite{Birkby18}. This technique requires high-resolution data in order to resolve individual absorption features from an atmospheric species, which are then cross-correlated with atmospheric models. The precision of the method increases with the number of lines included in the cross-correlation \citep[e.g.][]{Birkby18}. That means molecules (which have rotation-vibration transitions that produce thousands of absorption lines) are ideal targets, and are thus the focus of our analysis.

\begin{figure*}
    \centering
    \includegraphics{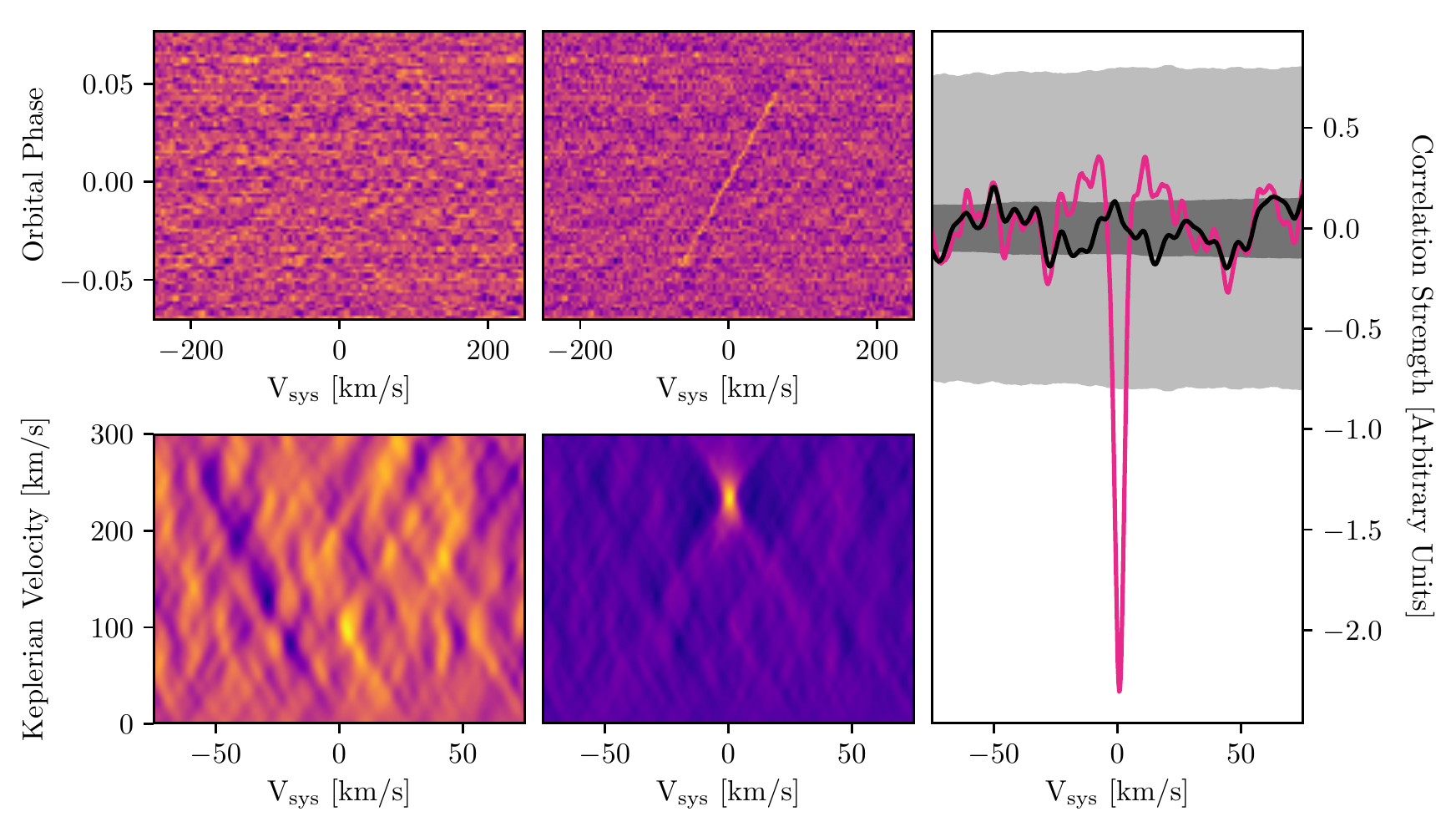}
    \caption{An example of the data analysis routine used in this work. The top two panels on the left show the results of carrying out the Doppler cross-correlation process as described in Section \ref{subsec:Doppler}. The top-left panel shows the results of cross-correlating one order of the first night of SPIRou data (N${}_7$) with a model containing HCN at a volume mixing ratio of 0.1\% and a mean molecular weight of 2 amu. The top-middle panel shows the same process, but with the atmospheric model injected into the data at 10 times its nominal strength (see Section \ref{subsec:injection} for a description of the model injection process). The planetary signal is visible as a diagonal line corresponding to the changing radial velocity of the planet, {which varies from $\sim -60$ km/s to $\sim +60$ km/s over the course of the transit}. The bottom two panels on the left show the results of phase-folding the corresponding top panels, as described in Section \ref{subsec:Doppler}, with the data-only phase-folded plot on the left and the data-plus-model phase-folded plot on the right. The panel on the right shows a horizontal slice of the phase-folded plots at the Keplerian velocity of 55~Cnc~e; the black line shows the data-only slice while the magenta line shows the data-plus-model slice. The dark- and light-grey contours show 1$\sigma$ and 3$\sigma$ confidence levels, respectively; the process used to generate these is described in Section \ref{subsec:significance}. We see that the data-only slice (black line) does not exceed 3$\sigma$ whereas the data-plus-model slice (magenta line) does, indicating that if this model were present in the atmosphere of 55~Cnc~e, we would have been able to detect it at a confidence level of $>3\sigma$.}
    \label{fig:method}
\end{figure*}

Following \cite{Snellen10} and \cite{Deibert19}, among others, we Doppler-shifted our atmospheric models to a range of velocities, and cross-correlated these with each frame of the data. We then phase-folded the correlations from the in-transit frames to a range of systemic velocities and summed over each velocity in order to obtain a grid of correlation strengths as a function of both systemic and Keplerian velocities. The grids were then summed over all relevant orders and all nights of our observations to increase the detection strength. An example of the analysis is shown in Fig. \ref{fig:method}, and the method is described in further detail in Section \ref{subsec:Doppler}.

Note that we have only included orders for which the atmospheric model in question contains absorption features. Examples of these atmospheric models are displayed in Appendix \ref{app:models}, and the models are described in further detail in the next section.

\subsection{Atmospheric Models}
\label{subsec:models}
{Our analysis involves independently testing for the presence of various molecules (HCN, NH${}_3$, C${}_2$H${}_2$, CO, CO${}_2$, and H${}_2$O) in the atmosphere of 55 Cnc e through cross-correlation with so-called ``parametric models'' similar to those used in \cite{Jindal20}. We explore a range of volume mixing ratios (VMRs) and mean molecular weights ($\mu$'s) of each of these molecules, described in further detail below. We note that these models are not self-consistent, and are instead meant to provide us with an initial insight into the exoplanet's atmosphere and the limitations of our method. Furthermore, various models (i.e. those with low mean molecular weights; see the top-left panels of the figures in the following section) are non-physical. Such models allow us to test our method, as they are easily-recoverable by the model injection/recovery process (see Section \ref{subsec:injection}). We describe the calculations of all models below.}

\subsubsection{Atmospheric Model Calculation}
\label{subsubsec:modelcalc}

To constrain the volume mixing ratios (VMRs) and mean molecular weights ($\mu$'s) of HCN, NH${}_3$, C${}_2$H${}_2$, CO, CO${}_2$, and H${}_2$O in the atmosphere of 55~Cnc~e, we generated a set of models with varying VMRs and $\mu$'s for each of these compounds individually embedded in an inert H$_{\rm{2}}$ atmosphere. The code used to generate these models is an updated version of the {line-by-line, plane parallel radiative transfer code} used in, e.g., \cite{Esteves17,Jindal20}.

Each model was generated on a wavelength grid spanning the full range of our observations, i.e. between $\sim$~950 and 2500 nm with a velocity step-size of 1 km/s. 
The model atmosphere was calculated across 50 atmospheric layers, and {opacities were integrated along slanted paths from the direction of the star to the observer}. Each model includes only a single molecule, {and we assume a Voigt profile for the lines with a line wing cutoff of 100/cm}. {The models were temperature-broadened using standard database parameters and included pressure-broadening coefficients from HITRAN \citep{Gordon17}. The line strengths were adjusted for temperature at each layer.} We assume that the VMR is constant throughout the atmosphere. In addition to molecular absorption, Rayleigh scattering and H$_2$-H$_2$ collision-induced absorption were also taken into account in all of our models \cite[][]{Borysow01, Borysow02}. As in \cite{Esteves17} and \cite{Jindal20}, we account for the geometry during transit when performing the radiative transfer. Like \cite{Jindal20}, we adjust the radius at the bottom of the models iteratively in order to match, on average, the measured value of $R_p/R_*$.

In the case of HCN, C${}_2$H${}_2$, and NH${}_3$, we made use of the HITRAN line lists (\citealt{Gordon17}). For CO, CO${}_2$, and H${}_2$O, we made use of the HITEMP line lists (\citealt{Rothman10}). {As in \cite{Jindal20}, we make use of the full line lists for our models.} {We note that as some of these line lists are incomplete (for e.g., the C${}_2$H${}_2$ line list), the shapes of various molecular bands in our models are not completely accurate: for example, several of the models presented in Appendix \ref{app:models} show sharp cutoffs at both ends of their molecular bands.}

To compare these models with our data, we convolved each of them to the resolution of our observations using a Gaussian kernel. We then spline-interpolated to the same wavelength grid when calculating the cross-correlation function.

Example models for each molecule are displayed in Appendix \ref{app:models}.

\subsection{Doppler Cross-Correlation}
\label{subsec:Doppler}
For each night of observations, our data were cross-correlated with models at Doppler shifts spanning $-250$~km/s to $250$~km/s, with 1 km/s steps between each velocity. This is shown in the top-left panels of Fig. \ref{fig:method}. Next, we phase-folded the correlation signal from the in-transit frames by shifting each correlation to the reference frame of 55~Cnc~e. The in-transit frames were determined using a model light curve generated with the \texttt{occultquad} package from \cite{Mandel2002}. The parameters used for this model light curve, including limb-darkening parameters from \cite{Claret2004}, are summarized in Table \ref{tab:parameters}. In order to account for the fact that the planetary radial velocity is not known with perfect precision, as well as to better understand the behavior of the cross-correlation function at velocities outside of the planetary rest frame (e.g. to investigate spurious correlation peaks due to residual telluric lines), we created a correlation map over a wide range of systemic velocities (V${}_\mathrm{sys}$) and planetary orbital velocities ($K_p \sin 2 \pi \phi(t)$, where $\phi(t)$ is the orbital phase). We created the map for planetary RV semi-amplitude ($K_p$) values ranging from 1 to 300 km/s, with a 0.5 km/s step between each value. An example is shown in the bottom-left panels of Fig. \ref{fig:method}.

For each model, these correlation grids were summed over orders containing significant molecular absorption, and then summed over all observations.

\subsection{Model Injection/Recovery Tests}
\label{subsec:injection}
To assess the robustness of our pipeline, and to place constraints on the atmospheric composition of 55~Cnc~e, we performed injection and recovery tests for each model used in our analysis. {The goal of these tests was to determine whether or not our data reduction process, including the \textsc{sysrem} algorithm, removes any signal that may be present, as well as to place sensitivity limits on our results}.

We {carried out these injection/recovery tests} by multiplying the in-transit frames of each observation (prior to any data reduction past initial processing at the telescope) by an atmospheric model shifted to the frame of the exoplanet {to create a synthetic ``model + data'' data set}. Note that the model was only added to the in-transit frames of our data. This was done for each model described in Section \ref{subsec:models}, each order of every observation, and each night of observation. The 
{synthetic ``model + data'' spectra }were then processed through our data reduction pipeline described in Section \ref{sec:reduc}, and analyzed using the Doppler cross-correlation method described in Section \ref{subsec:Doppler}. An example of the process is shown in the middle two panels of Fig. \ref{fig:method}.

As a result, we were able to assess {the sensitivity limits of our analysis, by determining which synthetic spectra were recovered by our pipeline. For example, the top-left panel in Fig. \ref{fig:hcn_result} shows a case where a correlation between a model atmosphere and the corresponding synthetic data set resulted in a detection (magenta line), whereas a correlation between that same model atmosphere and the true data set (black line) did not result in a detection. Likewise, the top-right panel in Fig. \ref{fig:hcn_result} shows a case where neither a correlation between a model atmosphere and the corresponding synthetic data set (magenta line) or the correlation between that same model atmosphere and the true data set (black line) resulted in a detection. The model atmosphere in question (HCN with a VMR of 0.1\% and a mean molecular weight of 10 amu) was beyond the sensitivity limits of our detection pipeline.}

{We note that various models with non-physical, low mean molecular weights (e.g. the top-left panel in Figure \ref{fig:h2o_result}) allow us to confirm that the model injection/recovery process is working as expected. Such models are easily recovered by the injection/recovery process, as seen in the top-left panels of the figures in the following section.}

{This model injection/recovery process thus} allowed us to place constraints on both the atmospheric makeup of the exoplanet as well as the detection limits of our observations.

\subsection{Detection Significance}
\label{subsec:significance}
In order to assess the significance of our results, we followed the methods described in \cite{Esteves17} and \cite{Deibert19}. We derived 1$\sigma$ and 3$\sigma$ confidence levels for each feature in each night of the data by randomly selecting a set of {out-of-transit} frames corresponding to the number of in-transit frames for each night, assigning an in-transit phase to each of these spectra, and following through with the cross-correlation and phase-folding process as described in Section \ref{subsec:Doppler}. We repeated this process 10,000 times, sorted the data, and then selected the 1$\sigma$ and 3$\sigma$ confidence levels based on the outcome. These levels were then compared to the correlation strengths of each model in order to assess their significance. An example is shown in the rightmost panel of Fig. \ref{fig:method}.

\section{Results}
\label{sec:discussion}
In this section we present the results of applying the Doppler cross-correlation method, as described in Section \ref{subsec:Doppler}, to our observations. As noted previously, we are searching for absorption features caused by the presence of atmospheric HCN,  NH${}_3$, C${}_2$H${}_2$, CO, CO${}_2$, and H${}_2$O. Note that the figures in the following sections show only a subset of the models we analyzed, and all additional models not displayed in the figures did not result in significant detections or limits. {Note as well that we are probing a grid of mean molecular weights and VMRs, and various combinations result in models that are non-physical. For example, a high ($\sim 10$ \%) VMR for CO${}_2$ would result in a mean molecular weight higher than 2 amu. However, such models allow us to test our method and ensure that the injection/recovery process is working as expected.}

\subsection{Hydrogen Cyanide}
\begin{figure*}
    \centering
    \includegraphics{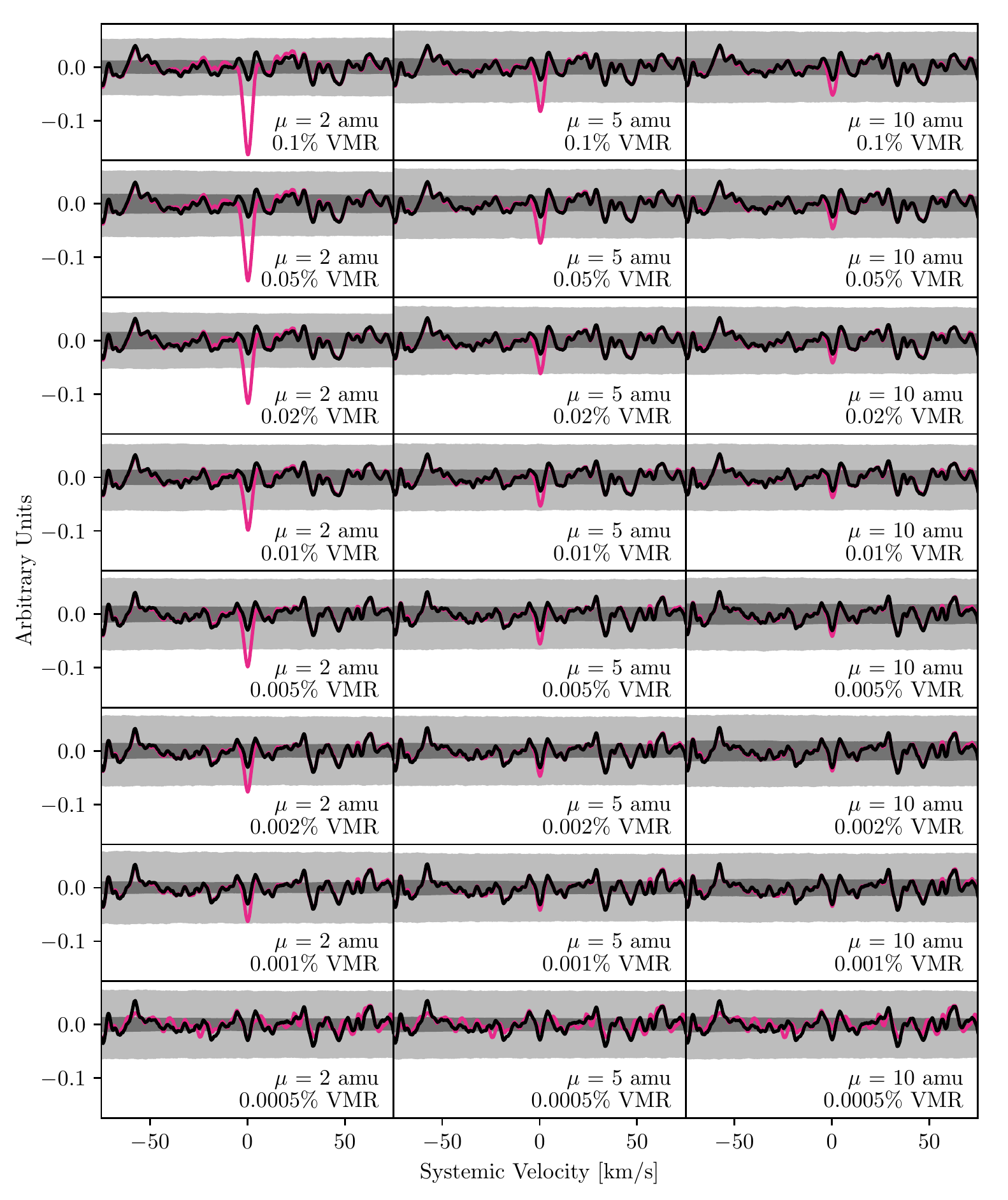}
    \caption{Results of injecting HCN models of various strengths into our data, and repeating the Doppler cross-correlation process. 
     In all panels, the black line represents the original data and the magenta line represents the data with an atmospheric model injected (see Section \ref{subsec:injection}). The volume mixing ratio and mean molecular weight of each model are indicated in the bottom right of the panel. The dark- and light-grey contours in each panel correspond to 1$\sigma$ and 3$\sigma$ confidence levels, respectively, and were calculated using the process described in Section \ref{subsec:significance}. The data have been phase-folded and sliced at the orbital velocity of 55~Cnc~e, $K_p \approx 231.4$ km/s. 
    We are able to rule out atmospheric HCN at a mean molecular weight of 2 amu with a volume mixing ratio as low as 0.001\%, and at a mean molecular weight of 5 amu with a volume mixing ratio as low as 0.02\%. We note that any additional models which were analyzed but not displayed in this figure did not yield significant detections or limits.}
    \label{fig:hcn_result}
\end{figure*}

Fig. \ref{fig:hcn_result} shows the results of our search for HCN in the atmosphere of 55~Cnc~e. Our aim was to further investigate the results of \cite{Tsiaras16}, who used HST/WFC3 observations to report the detection of an atmosphere and suggested that HCN is the most likely molecule able to account for the observed absorption features.

We analyzed a range of atmospheric models varying in volume mixing ratio from 0.1\% down to 5$\times 10^{-5}$\%, and in mean molecular weight from 2 to 20 amu. Our model spectra were generated using HITRAN2016 (\citealt{Gordon17}). We note that the results depend on the accuracy of the line list used, thus future studies with updated line lists may yield different outcomes.

As seen in Fig. \ref{fig:hcn_result}, we can rule out the presence of HCN in the atmosphere of 55~Cnc~e at a mean molecular weight of 2 amu with a volume mixing ratio of 0.001\%; if the mean molecular weight is increased to 5 amu, the lowest volume mixing ratio that we can rule out is 0.02\%. 

\begin{figure*}
    \centering
    \includegraphics{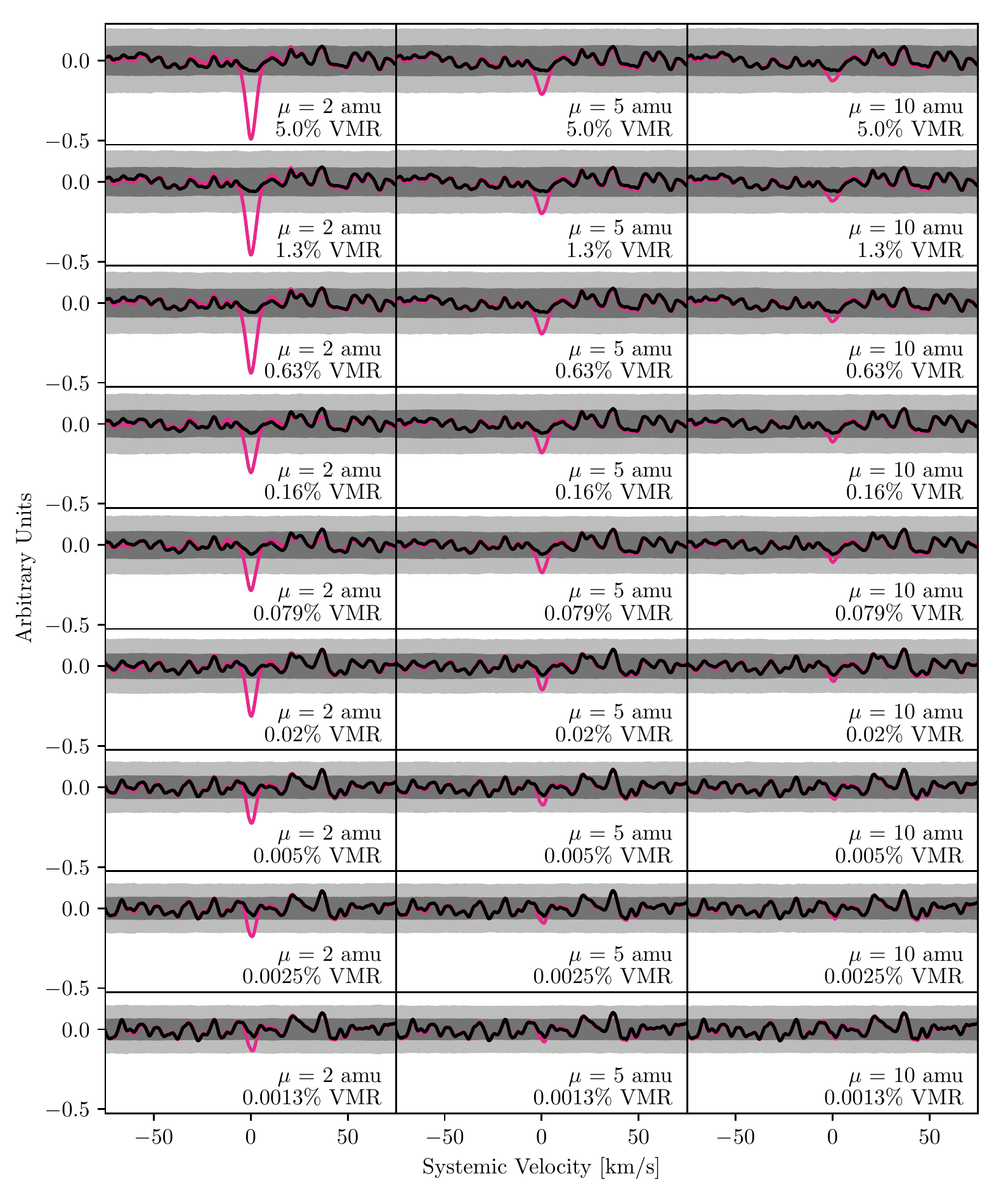}
    \caption{Results of injecting NH${}_3$ models of various strengths into our data, and repeating the Doppler cross-correlation process. The panels are as described in the caption of Fig. \ref{fig:hcn_result}. We are able to rule out the presence of NH${}_3$ in the atmosphere of 55~Cnc~e at a mean molecular weight of 2 amu with a volume mixing ratio as low as 0.0025\%, and at a mean molecular weight of 5 amu with a volume mixing ratio as low as 0.08\%. We note that any additional models with lower VMRs or higher mean molecular weights which were analyzed but not displayed in this figure did not yield significant detections or limits.}
    \label{fig:nh3_result}
\end{figure*}

\subsection{Ammonia}
Fig. \ref{fig:nh3_result} shows the results of our search for NH${}_3$ in the atmosphere of 55~Cnc~e. 

We analyzed a set of models ranging in volume mixing ratio from 5\% to $10^{-8}$\%, and mean molecular weight ranging from 2 amu to 20 amu.

As can be seen in Fig. \ref{fig:nh3_result}, we are able to rule NH${}_3$ out of the atmosphere at a mean molecular weight of 2 amu with a volume mixing ratio as low as 0.0025\%; if the mean molecular weight is increased to 5 amu, we can still rule NH${}_3$ out with a volume mixing ratio as low as 0.08\%. 

\subsection{Acetylene}
\begin{figure*}
    \centering
    \includegraphics{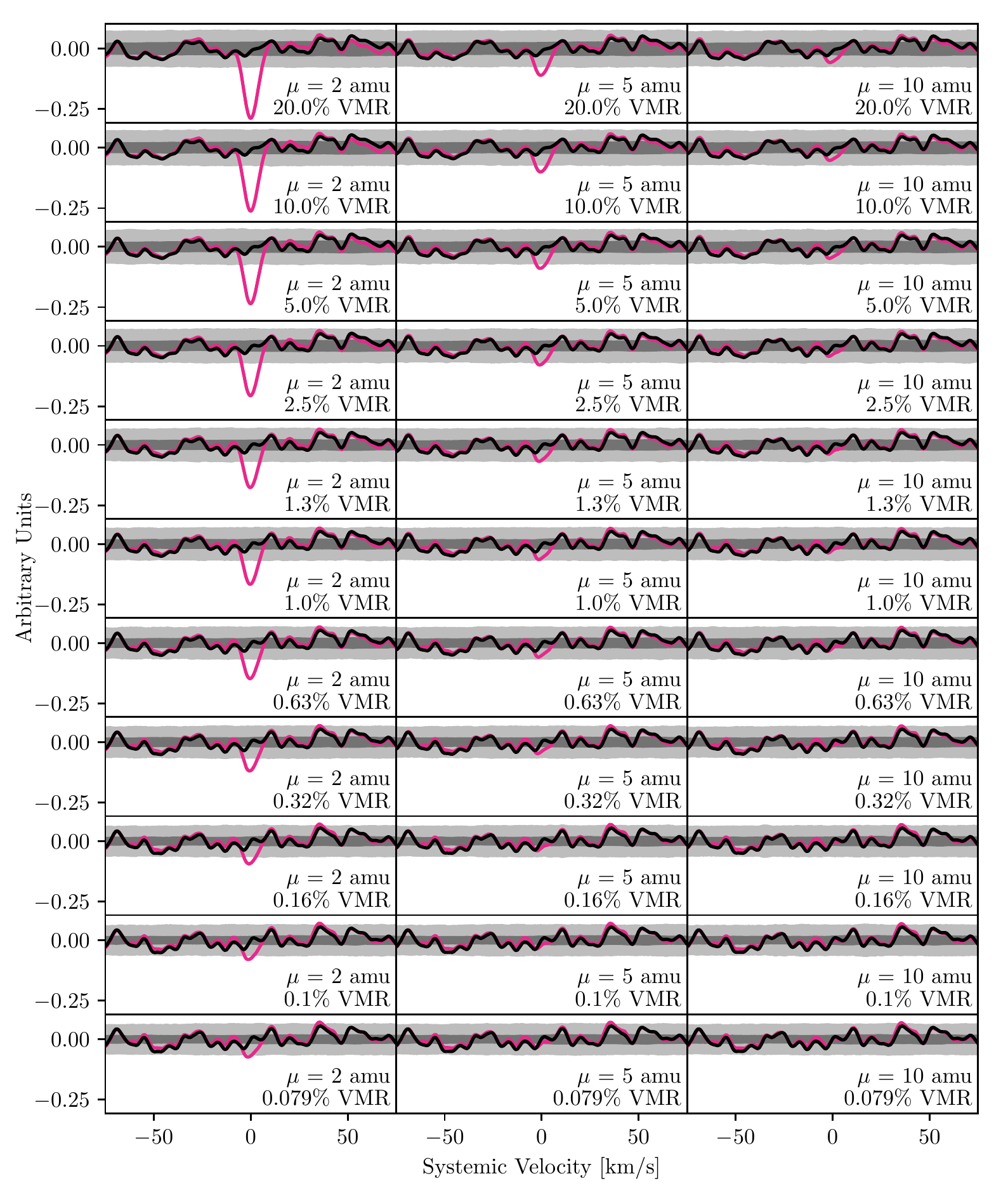}
    \caption{Results of injecting C${}_2$H${}_2$ models of various strengths into our data, and repeating the Doppler cross-correlation process. The panels are as described in the caption of Fig. \ref{fig:hcn_result}. We are able to rule C${}_2$H${}_2$ out of the atmosphere of 55~Cnc~e at a mean molecular weight of 2 amu with a volume mixing ratio as low as 0.08\%, and at a mean molecular weight of 5 amu with a volume mixing ratio as low as 1.0\%. We note that any additional models with lower VMRs or higher mean molecular weights which were analyzed but not displayed in this figure did not yield significant detections or limits.}
    \label{fig:c2h2_result}
\end{figure*}

Fig. \ref{fig:c2h2_result} shows the results of our search for C${}_2$H${}_2$ in the atmosphere of 55~Cnc~e.

We analyzed a set of models ranging in volume mixing ratio from 20\% to $10^{-8}$\%, and mean molecular weight ranging from 2 amu to 20 amu.

As can be seen in Fig. \ref{fig:c2h2_result}, we are able to rule C${}_2$H${}_2$ out of the atmosphere of 55~Cnc~e at a mean molecular weight of 2 amu with a volume mixing ratio as low as 0.08\%, and at a mean molecular weight of 5 amu with a volume mixing ratio as low as 1.0\%. 

\subsection{Carbon Monoxide}
\label{subsec:co}
\begin{figure*}
    \centering
    \includegraphics{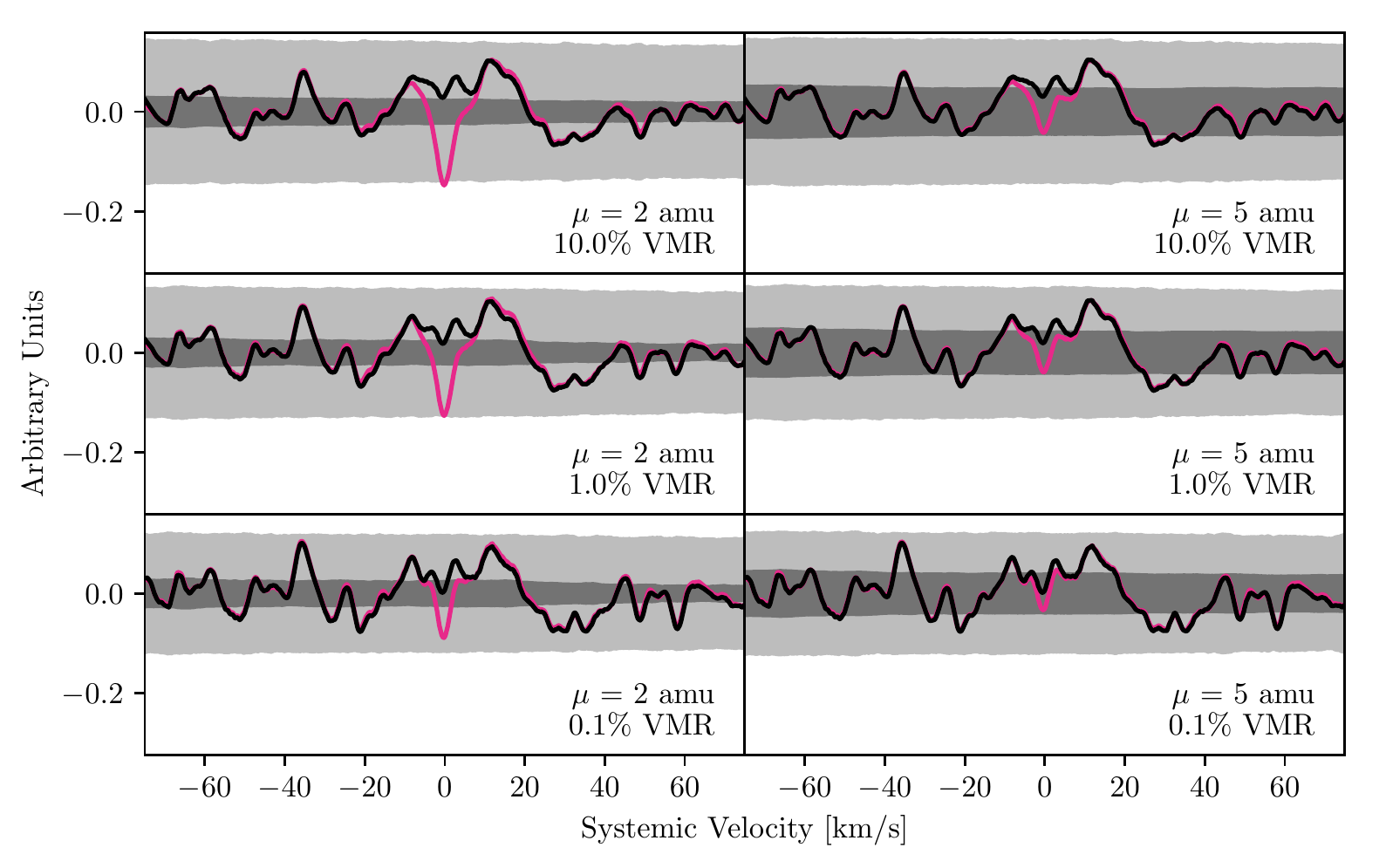}
    \caption{Results of injecting CO models of various strengths into our data, and repeating the Doppler cross-correlation process.
     The panels are as described in the caption of Fig. \ref{fig:hcn_result}.
    We are able to tentatively rule out the presence of CO in the atmosphere of 55~Cnc~e at a mean molecular weight of 2 amu down to a volume mixing ratio of 1.0\% at a confidence of 3$\sigma$, as can be seen in the top two panels on the left. However, we caution that there are many additional peaks surpassing 1$\sigma$ in the results (likely due to noise in the data) and that this may have affected our model injection/recovery process.}
    \label{fig:co}
\end{figure*}

The results of our search for CO in the atmosphere of 55~Cnc~e are summarized in Fig. \ref{fig:co}.

We analyzed a range of atmospheric models varying in volume mixing ratio from 10\% down to $10^{-6}$\%, and in mean molecular weight from 2 to 30 amu. 

As shown in Fig. \ref{fig:co}, we are unable to detect atmospheric CO in our data. However, we are able to tentatively rule out the possibility of CO being present in the atmosphere of 55~Cnc~e at a mean molecular weight of 2 amu and a volume mixing ratio as low as 1.0\% at the 3$\sigma$ level. Note however that there are numerous additional features with peaks $>1\sigma$ that are likely due to noise in the data (see, e.g., the discussion in \citealt{Esteves17}, who noticed a similar phenomenon for optical transit data of 55~Cnc~e). For this reason, we caution that our limits on the presence of CO are only tentative, and warrant further investigation.

\begin{figure*}
    \centering
    \includegraphics{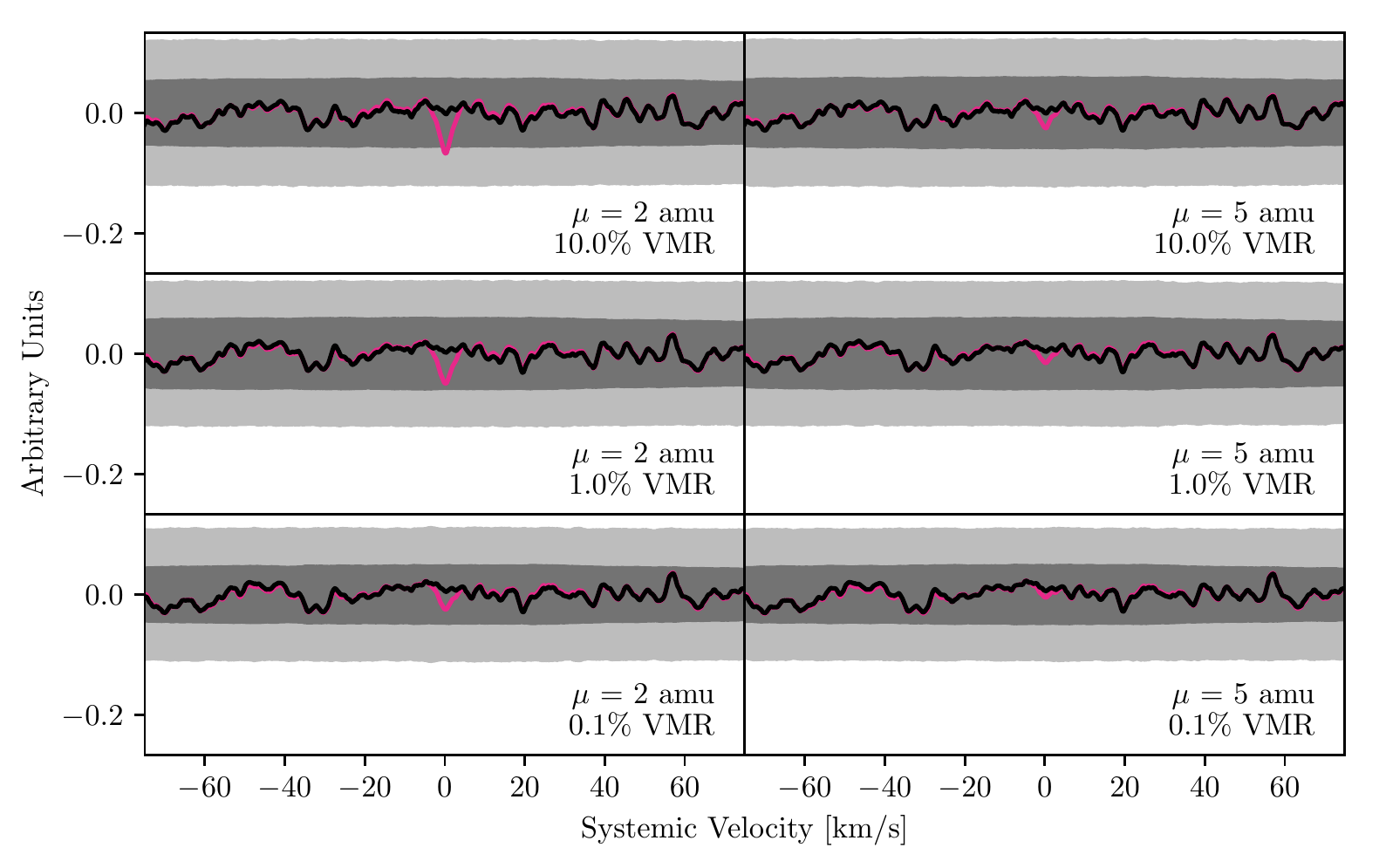}
    \caption{Results of injecting CO${}_2$ models of various strengths into our data, and repeating the Doppler cross-correlation process. The figures are as described in the caption of Fig. \ref{fig:hcn_result}. As neither the data (black lines) nor the data with models injected (magenta lines) surpass the 3$\sigma$ confidence levels, we have neither detected nor can place limits of the presence of CO${}_2$ in the atmosphere of 55~Cnc~e.}
    \label{fig:co2_result}
\end{figure*}

\subsection{Carbon Dioxide}
\label{subsec:co2}
The results of our search for CO${}_2$ in the atmosphere of 55~Cnc~e are summarized in Fig. \ref{fig:co2_result}.

As was the case with CO (see Section \ref{subsec:co}), we analyzed a range of atmospheric models varying in volume mixing ratio from 10\% down to $10^{-6}$\%, and in mean molecular weight from 2 to 30 amu.

As can be seen in Fig. \ref{fig:co2_result}, we are neither able to detect atmospheric CO${}_2$ in our data nor place any significant constraints on its presence. We note that while a peak is present in the injected data at a systemic velocity of 0 km/s and the orbital velocity of 55~Cnc~e, as would be expected for a detection, it does not surpass 3$\sigma$ and therefore is not sufficiently significant to make any conclusions about the presence or lack of atmospheric CO${}_2$.

\subsection{Water}
\label{subsec:h2o}
\begin{figure*}
    \centering
    \includegraphics{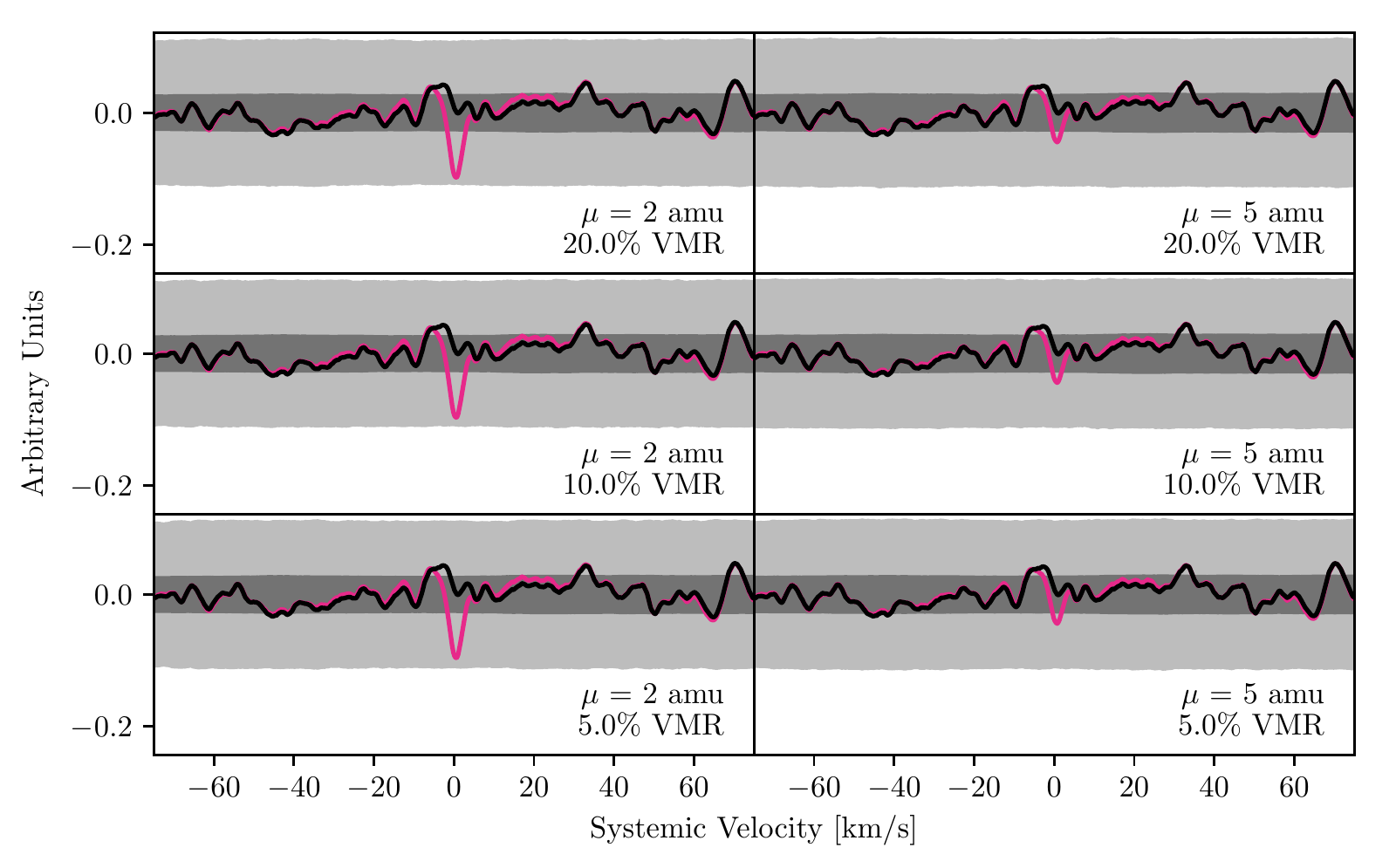}
    \caption{Results of injecting H${}_2$O models of various strengths into our data, and repeating the Doppler cross-correlation process. The figures are as described in the caption of Fig. \ref{fig:hcn_result}. As neither the data (black lines) nor the data with models injected (magenta lines) surpass the 3$\sigma$ confidence levels, we have neither detected nor can place limits of the presence of H${}_2$O in the atmosphere of 55~Cnc~e based on these data. We do, however, note the presence of a peak at the expected location that approaches 3$\sigma$ for each model displayed with a mean molecular weight of 2 amu, indicating that we may be able to rule these models out with additional observations.}
    \label{fig:h2o_result}
\end{figure*}

The results of our search for H${}_2$O in the atmosphere of 55~Cnc~e are summarized in Fig. \ref{fig:h2o_result}.

We analyzed a range of atmospheric models varying in volume mixing ratio from 20\% down to $10^{-8}$\%, and in mean molecular weight from 2 to 20 amu. 

As was the case with CO${}_2$ (see Section \ref{subsec:co2}), none of the injected models seen in Fig. \ref{fig:h2o_result} surpass the 3$\sigma$ confidence level, and thus we are unable to place any limits of the presence of H${}_2$O in the atmosphere of 55~Cnc~e with these data alone. This is likely due to the imperfect removal of telluric water lines in the data (see Section \ref{subsec:sysrem} and the additional discussion in Section \ref{subsec:comparison} and Appendix \ref{app:tests} for further details).

\section{Discussion} \label{sec:actual_discussion}

\subsection{Comparison with \citealt{Tsiaras16}}
Our model injection/recovery tests for HCN (see Fig.~\ref{fig:hcn_result}) significantly narrow the parameter space of potential atmospheres that are consistent with the results reported in \cite{Tsiaras16}. 
These authors sampled volume mixing ratios from 1 to $10^{-8}$ and mean molecular weights from 2 to 10 amu. Their analysis indicates that the mean molecular weight of the atmosphere peaks at roughly 4 amu, with higher values being unlikely. Furthermore, strong absorption seen near 1.4 and 1.6 $\mu$m in their analysis indicates that the mean molecular weight is relatively low. They also reported that (of the molecules considered in their fit) HCN is the most likely absorber able to explain their observed absorption features, and that a scenario with a high volume mixing ratio for HCN is favoured, though values as low as 0.001\% are acceptable.

While our results do rule out the most likely scenarios considered by \cite{Tsiaras16}, they are not in complete disagreement. A mean molecular weight of 2 amu at volume mixing ratios consistent with the \cite{Tsiaras16} analysis is ruled out by our observations; however, it is possible that the mean molecular weight is slightly greater ($\sim$4 or 5 amu), and that the volume mixing ratio is between 0.02\% and 0.001\%.
If the detection in \citealt{Tsiaras16} is real, a higher mean-molecular weight would be unlikely, since that would significantly mute the features in the transmission spectrum.

While additional observations will be necessary to determine whether or not a low volume mixing ratio/high mean molecular weight scenario can be ruled out completely, our analysis has provided significantly improved constraints on the presence of HCN and vastly decreased the parameter space of possible atmospheres. Our analysis has also provided constraints on the presence of NH${}_3$ and C${}_2$H${}_2$, and suggests that if any of these three molecules are indeed present in the atmosphere of 55~Cnc~e, they are likely to have high mean molecular weights, low volume mixing ratios, or both. We note that previous observations \citep[e.g.][]{Ehrenreich12,Demory16} do support a high-mean-molecular-weight atmosphere. 

Finally, we caution that these constraints rely on the accuracy of the line lists used to generate our models, and future updates to these line lists could yield different results. The issue has been discussed in detail in a number of previous studies. In the case of TiO, for example, \cite{Hoeijmakers15} demonstrated that inaccurate line lists can hamper high-resolution retrievals. Likewise, \cite{Webb20} observed absorption due to H${}_2$O in high-resolution VLT/CRIRES spectra of HD 179949 b, but found a weak dependence on the line list used for the cross-correlation. Updates to the available line lists for each of the molecules used in this analysis could similarly impact our detection capabilities, and we therefore note that the particulars of line lists should be kept in mind when interpreting our results.

\subsection{Comparison with High-Resolution Optical Spectroscopy}
\label{subsec:comparison}
We note that previous studies \citep{Esteves17,Jindal20} have been able to place significant constraints on the presence of H${}_2$O in the atmosphere of 55~Cnc~e using high-resolution spectra at optical wavelengths. In particular, recent work by \cite{Jindal20} resulted in a 3$\sigma$ lower limit of 15 amu on the mean molecular weight of 55~Cnc~e's atmosphere, assuming it is water-rich (i.e. has a volume mixing ratio of $>0.1$\%) and that it does not have thick clouds. Our observations do not improve on these constraints.

Given the fact that the NIR wavelength ranges of CARMENES and SPIRou contain more absorption lines due to H${}_2$O than the optical wavelength ranges of these previous works (e.g. compare our Fig. \ref{fig:h2o_examplemodel} with Fig. 8 of \citealt{Esteves17} and Fig. 2 of \citealt{Jindal20}), it could be expected that our data (which cover the same number of transits as \citealt{Jindal20}) should give the same or even better constraints than those analyses. To investigate the discrepancy, we carried out a number of tests designed to probe the limits of our technique in the context of NIR H${}_2$O (and CO${}_2$) absorption. The full details of these tests are presented in Appendix \ref{app:tests}. 

 We conclude that the discrepancies between this work and \cite{Esteves17} and \cite{Jindal20} are largely due to our inability to remove telluric water features fully from our data. While the overall wavelength coverage of the observations used in this work is indeed broader than that of \cite{Jindal20}, and therefore contains a greater number of absorption lines due to H${}_2$O, the telluric contamination at these redder wavelengths is far more severe than at the wavelengths considered in \cite{Jindal20}. As the individual absorption lines due to the exoplanet's atmosphere are weak, our method relies heavily on our ability to combine the signals from many absorption lines, which in turn relies on our ability to remove telluric contamination adequately in the regions where these lines are present. \textsc{sysrem} is less efficient in regions of strong telluric contamination; this can be seen in a comparison between the residuals after applying \textsc{sysrem} in \cite{Jindal20} and those in this work (see the plots in Appendix \ref{app:reduc}). While additional iterations of \textsc{sysrem} may aid in removing some residual telluric contamination, we show in Appendix \ref{app:tests} that additional iterations may also begin to remove the model itself, and thus will not improve our detection capabilities. We also note that the SNRs of our observations (see Table \ref{tab:obs}) are in some cases significantly lower than those of \cite{Jindal20}, which would limit our sensitivities further.

As mentioned in Section \ref{subsubsec:molec}, we also tried correcting for telluric absorption using Molecfit. We applied Molectit and Calctrans to every spectrum in each night of our CARMENES and SPIRou data. We then injected an atmospheric H${}_2$O model with a VMR of 20\% and a mean molecular weight of 2 amu (i.e. the model shown in Fig. \ref{fig:h2o_examplemodel}; the strongest water model of our sample) into the data and repeated this process. Then, we carried out the Doppler cross-correlation process as described in Section \ref{subsec:Doppler} in order to determine whether or not we were more readily able to recover an injected atmospheric H${}_2$O model with \textsc{sysrem} or with Molecfit. 

\begin{figure*}
    \centering
    \includegraphics{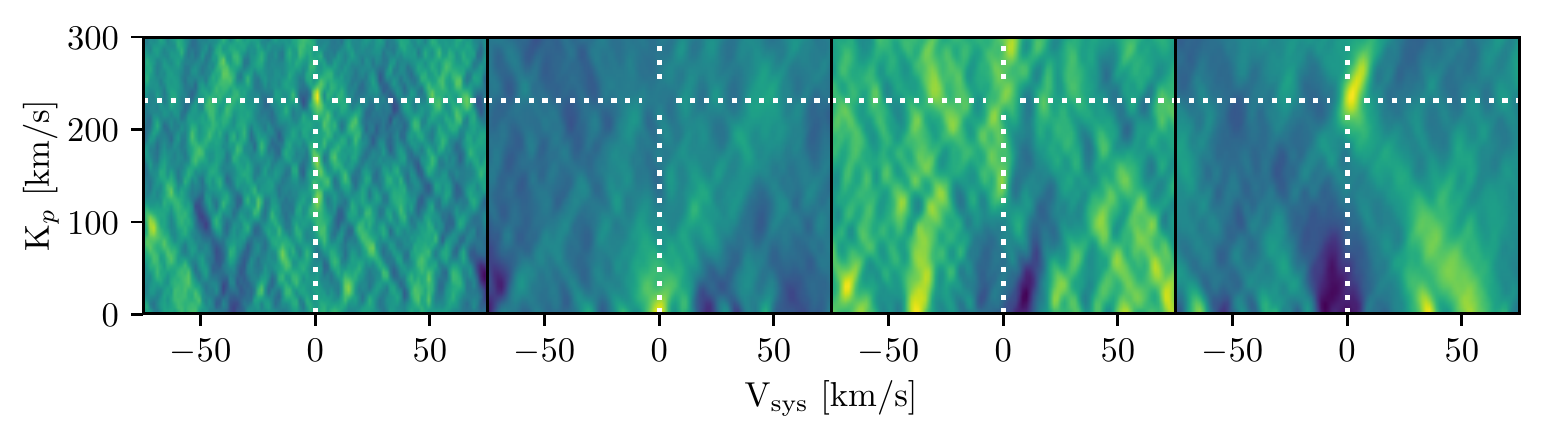}
    \caption{A comparison of our phase-folded CARMENES results for cases where telluric correction is carried out using \textsc{sysrem} (first panel) and Molecfit (second, third, and fourth panels). In all cases, we have injected an atmospheric H${}_2$O model with a VMR of 20\% and a mean molecular weight of 2 amu into our data. In the first panel we have corrected telluric absorption using \textsc{sysrem} (Section \ref{subsec:sysrem}); in the second panel we have corrected telluric absorption using Molecfit and corrected stellar absorption using a mean stellar line template; in the third panel we have corrected telluric absorption using Molecfit and corrected stellar absorption using 2 iterations of \textsc{sysrem}; and in the fourth panel we have injected the model at $100\times$ its nominal strength into a single night of data and corrected for telluric absorption using Molecfit and stellar absorption using 2 iterations of \textsc{sysrem}. In all cases, the correction is followed by the Doppler cross-correlation process as described in Section \ref{subsec:Doppler}. A peak is visible in the cross-correlation at the expected location for \textsc{sysrem} and Molecfit with a model injected at $100\times$ its nominal strength; in the cases where Molecfit is used on the model at $1\times$ its nominal strength, however, we are unable to detect a peak in the cross-correlation. The expected location is indicated by white dotted lines in all panels.}
    \label{fig:molecfit_vs_sysrem}
\end{figure*}

\begin{figure*}
    \centering
    \includegraphics{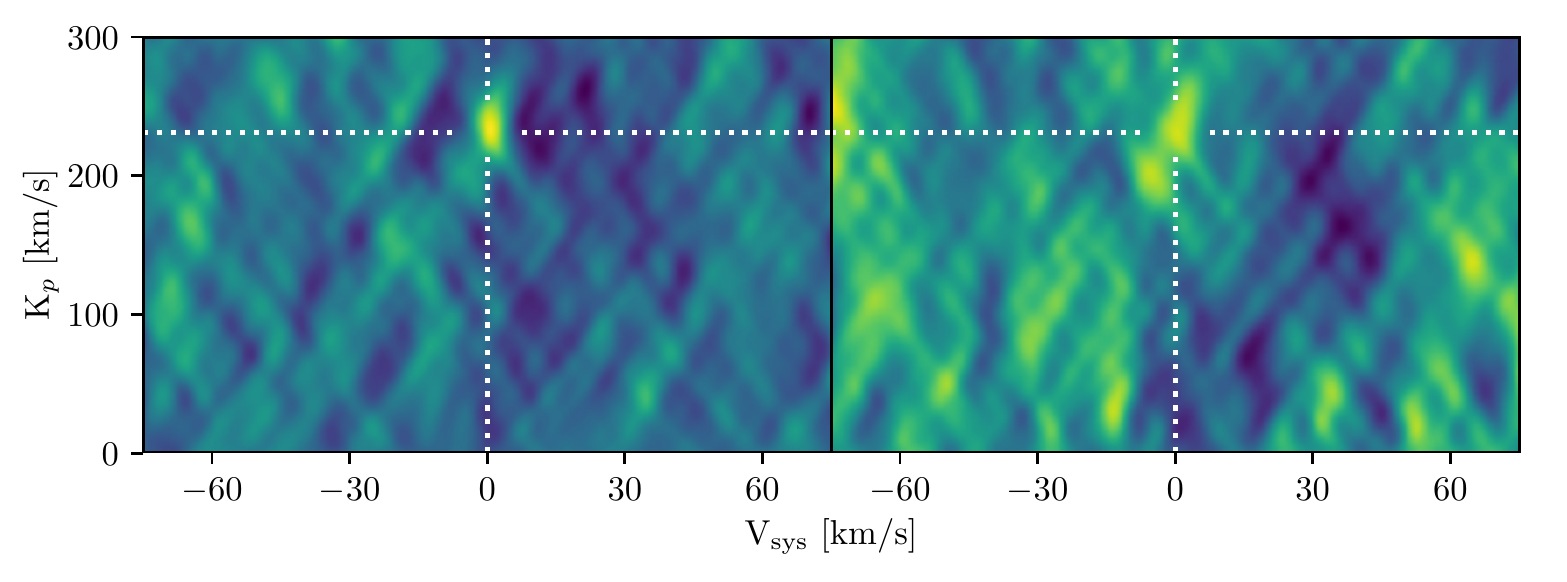}
    \caption{A comparison of our phase-folded SPRIou results for cases where telluric correction is carried out using \textsc{sysrem} (left) and Molecfit (right). In both bases, we have injected an atmospheric H${}_2$O model with a VMR of 20\% and a mean molecular weight of 2 amu. In the panel on the right, stellar absorption has been corrected with 2 iterations of \textsc{sysrem}. In this case we see that the injected model is recovered when the data are corrected with Molecfit; however, the recovered signal is not as strong as in the case where the data are corrected with \textsc{sysrem}.}
    \label{fig:molecfit_spirou}
\end{figure*}

We found that Molecfit does not improve our ability to detect an injected atmospheric H${}_2$O model. Instead, the data corrected with Molecfit yielded a weaker detection of the injected model than those corrected with \textsc{sysrem}. This is evident in Figs. \ref{fig:molecfit_vs_sysrem} and \ref{fig:molecfit_spirou}, where we compare the two for CARMENES and SPIRou observations, respectively. We tested two separate methods for removing stellar lines after applying Molecfit to our data: first, we subtracted a mean template of the stellar features remaining; and second, we applied two iterations of \textsc{sysrem} to the data after correcting them with Molecfit. In the case of CARMENES observations we are unable to recover the injected model (the middle two panels of Fig. \ref{fig:molecfit_vs_sysrem}); for SPIRou, on the other hand, we can recover the injected model at a lower significance than our original analysis (the second panel of Fig. \ref{fig:molecfit_spirou}).

As a check for our CARMENES observations, we also tested a case where we injected the model at 100 times its nominal strength into a subset of our data and followed the same reduction process using Molecfit. In this case, we were able to recover the injected model at the expected location.

We note that previous studies which have compared Molecfit and \textsc{sysrem} in the NIR have had similar results. In particular, \cite{SanchezLopez19} made a detection of water vapour in the atmosphere of HD 209458 b using CARMENES data that had been corrected with \textsc{sysrem}. After correcting their data with Molecfit, however, they were unable to recover the planetary signal. The authors noted that recent works have found the scatter in residuals for CRIRES data corrected with Molecfit to be 3 to 7\% \citep{UlmerMoll19}, and concluded that the fit uncertainties over the wavelength coverage of CARMENES are likely too large to detect the H${}_2$O features of the planet. Our analysis yields similar results: although we do not strongly detect the atmospheric H${}_2$O model with \textsc{sysrem}, there is a correlation peak at the expected location that approaches 3$\sigma$. When running Molefit on our CARMENES data, however, no such peak is present. In the case of SPIRou, a peak is visible at the expected location, but this signal is weaker than the case where \textsc{sysrem} was used instead (Fig. \ref{fig:molecfit_spirou}).

As a final test of what might be impacting our ability to recover injected H${}_2$O models, we followed \cite{Alonso19} and \cite{SanchezLopez19} in separately analyzing individual bands of our observations. We separated the CARMENES data into the Y, J, and H bands, and separated the SPIRou data into the Y, J, H, and K bands.

\begin{figure*}
    \centering
    \includegraphics[width=\textwidth]{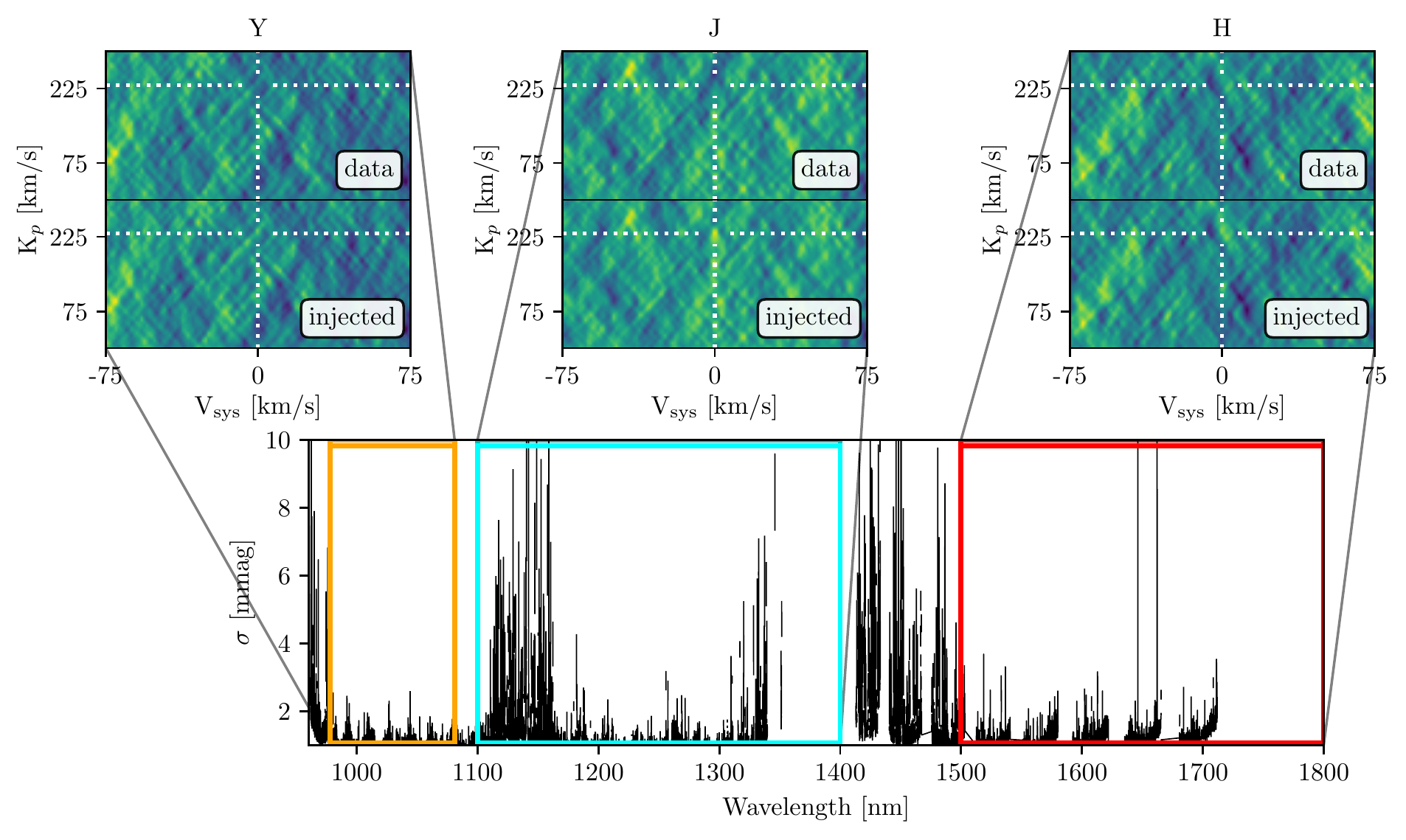}
    \caption{The results of our model injection/recovery process (see Section \ref{subsec:injection}) for the Y, J, and H bands of our CARMENES observations. These tests use a water model with a VMR of 20\% and a mean molecular weight of 2 amu (Fig. \ref{fig:h2o_examplemodel}). The primary plot shows $\sigma$ as a function of wavelength for the full wavelength range, and the insets show the phase-folded correlations for the data alone (top insets) and the data with a model injected (bottom insets). The insets are as described in the caption of Fig. \ref{fig:method}. The band is indicated in the title of each inset, and the corresponding region is marked in the primary plot. The white dotted lines show the expected location of the correlation signal. We see that the majority of the recovered signal comes from the J band.}
    \label{fig:carmenes-multiband}
\end{figure*}

\begin{figure*}
    \centering
    \includegraphics[width=\textwidth]{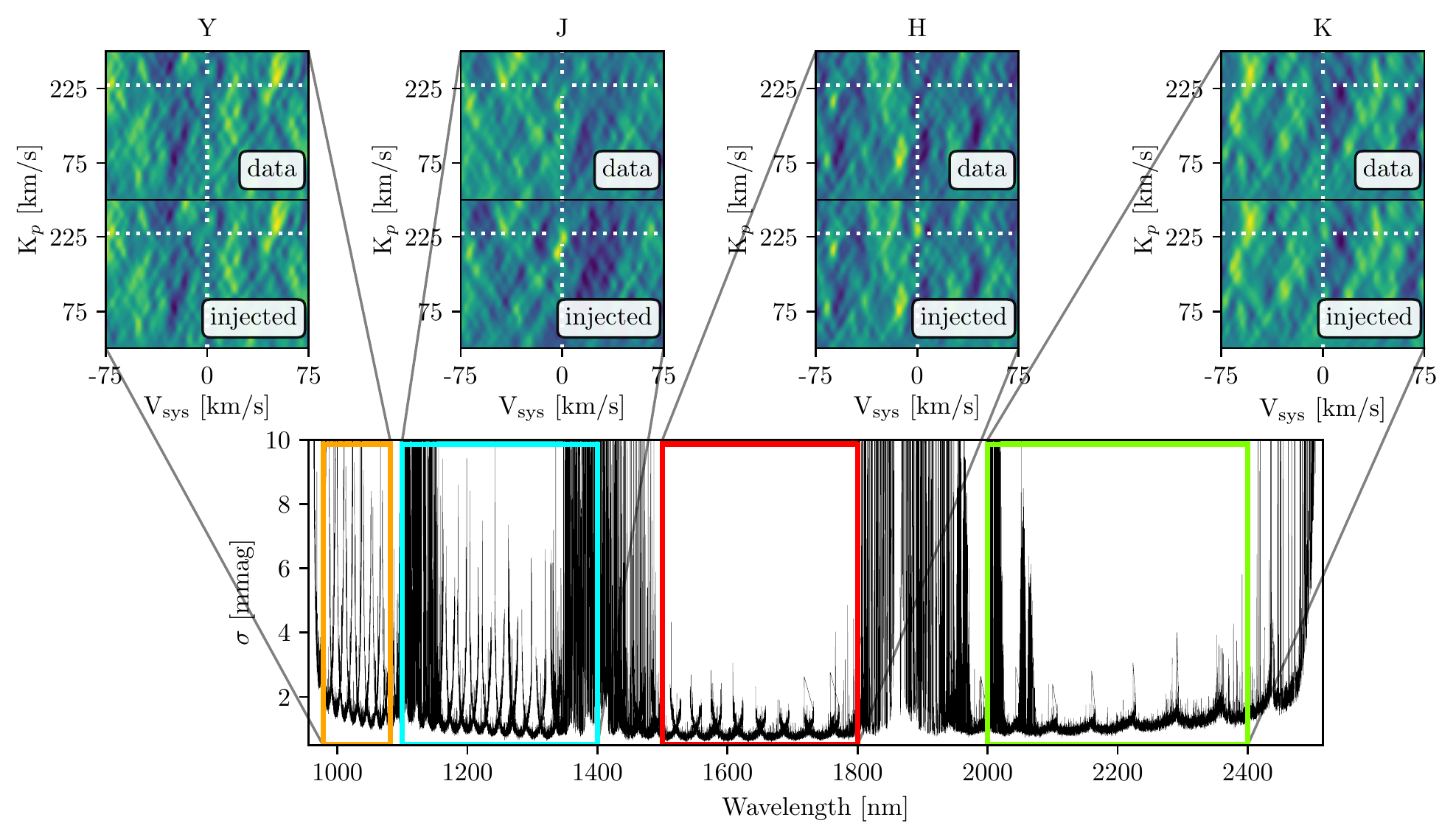}
    \caption{The same as Fig. \ref{fig:carmenes-multiband}, but for SPIRou. Here, the recovered signal comes largely from the J and H bands.}
    \label{fig:spirou-multiband}
\end{figure*}

Figs. \ref{fig:carmenes-multiband} and \ref{fig:spirou-multiband} show these results for CARMENES and SPIRou, respectively. In both cases the contribution to the signal comes almost exclusively from the J band (and, to a lesser extent, the H band). This suggests that although our observations span a wide wavelength range, only a small fraction of that wavelength range is contributing to our detection capabilities. The other water features present (see Fig. \ref{fig:h2o_examplemodel}) were likely not readily recovered by our model injection/recovery tests because they are comparatively weaker, and because \textsc{sysrem} was not able to remove telluric features completely in these regions (see the figures in Appendix \ref{app:reduc}, particularly the bottom two panels of each).

\begin{figure*}
    \centering
    \includegraphics{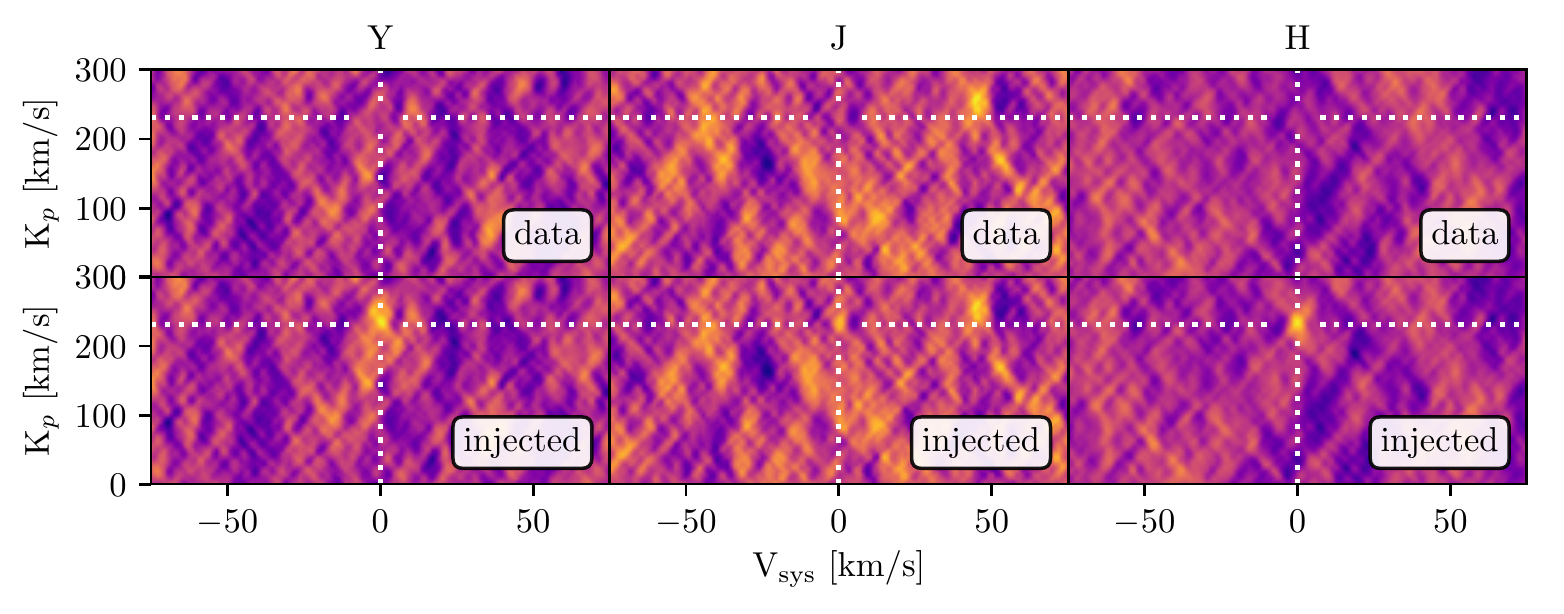}
    \caption{The same as the insets in Fig. \ref{fig:carmenes-multiband}, but for an HCN model with a VMR of 0.1\% and a mean molecular weight of 2~amu (Fig. \ref{fig:hcn_examplemodel}). We have omitted the plot of $\sigma$ as a function of wavelength here as it is the same as the primary plot in Fig. \ref{fig:carmenes-multiband}. We see that the majority of the recovered signal comes from the Y and H bands, but there is a peak present at the expected location in the J band as well.}
    \label{fig:hcn-multiband-carmenes}
\end{figure*}

\begin{figure*}
    \centering
    \includegraphics{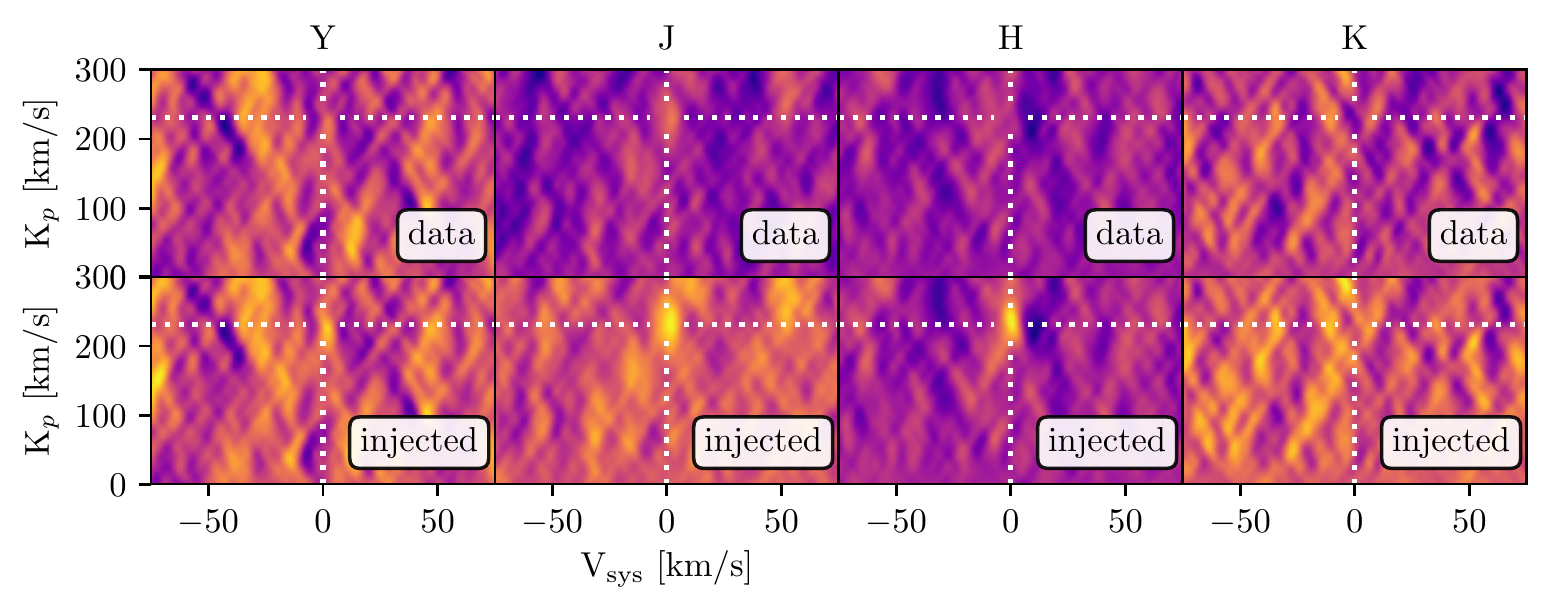}
    \caption{The same as the insets in Fig. \ref{fig:spirou-multiband}, but for an HCN model with a VMR of 0.1\% and a mean molecular weight of 2~amu (Fig. \ref{fig:hcn_examplemodel}). We see that the majority of the recovered signal comes from the J and H bands, with some contribution from the Y band; there is no significant correlation peak recovered in the K band.}
    \label{fig:hcn-multiband-spirou}
\end{figure*}

We also ran the individual band analysis on our strongest HCN model as a comparison. The results are shown in Figs. \ref{fig:hcn-multiband-carmenes} and \ref{fig:hcn-multiband-spirou} for CARMENES and SPIRou, respectively. In this case we see that a signal is recovered in all bands except the K band. For CARMENES the signal is strongest in the Y and H bands, whereas for SPIRou the signal is noticeably strongest in the J and H bands.

\subsection{A Nitrogen-Dominated Atmosphere?}
Several previous studies have pointed to a nitrogen-dominated atmosphere for 55~Cnc~e. \cite{Hammond17} used 3D calculations to show that the offset observed in the phase curve \citep{Demory16,Angelo17}, as well as the large day-night temperature contrast, may be explained by an N-dominated atmosphere. The analysis by \cite{Angelo17} also supports this scenario. To this end, \cite{Miguel19} explored the observable features in the spectra of an N-dominated atmosphere for 55~Cnc~e. Through analytical arguments, equilibrium chemistry calculations, and adopting Titan's elemental abundances as a potential composition, \cite{Miguel19} showed that although N${}_2$ is expected to be the most abundant molecule in the atmosphere (followed by H${}_2$ and CO), the transmission spectra should show strong features of NH${}_3$ and HCN. However, a decrease in the N/O ratio would tend to weaken these NH${}_3$ and HCN features.

The transmission spectra calculated in \cite{Miguel19} (their Fig. 3, right panel) range between 3 and 20 $\mu$m, and are thus outside the range of our own observations. Figs. 6 and 7 of \cite{Miguel19} show the mixing fraction of the most abundant observable molecules as a function of temperature at a pressure of $\sim$1.4 bar \citep[as calculated by][]{Angelo17}, and the mixing fractions as a function of pressure at different N/O ratios, respectively. For all temperatures considered by \cite{Miguel19} at a pressure of $\sim$~1.4 bar,  the volume mixing ratio of NH${}_3$ is less than $\sim10^{-4}$\%, while at all pressures considered and for all N/O ratios considered, the volume mixing ratio of NH${}_3$ is less than $\sim10^{-5}$\%. Our analysis rules out most low-mean-molecular-weight, high-volume-mixing-ratio scenarios, which is consistent with expected volume mixing ratios from \cite{Miguel19}, and with previous work that has combined mass and radius measurements with interior modeling to show that the atmosphere should have a high mean molecular weight \citep{demory11,winn11,Bourrier18}.

\subsection{A Cloudy Atmosphere?}
We note that the limits we have placed throughout this section assume that any signals present in the atmosphere are not obscured by clouds or hazes.
However, the presence of clouds or hazes could act to obscure atmospheric signals, limiting our ability to make detections. Therefore, our results may also be consistent with a cloudy atmosphere. The limits we have placed throughout this analysis are only relevant in the case of a cloud-free atmosphere, as the models were injected into the data under the assumption that they were not obscured at any altitude by clouds or hazes. {We note as well that various other investigations into the composition of 55 Cancri e's atmosphere \citep{Tsiaras16,Esteves17,Jindal20} made similar assumptions, meaning that those limits also pertain to the case in which 55 Cnc e does not harbour clouds.}

While an in-depth exploration of the effects of clouds or hazes on the atmosphere could prove insightful, such an analysis is beyond the scope of this work.
{However, we note that an analysis by \cite{Mahapatra17} found that despite 55 Cnc e's high equilibrium temperature ($\sim 2400$ K, \citealt{Demory16b}), conditions on 55 Cnc e may allow for mineral clouds to form. Such cloud formation is likely only possible in a thin atmospheric region, and requires strong vertical replenishment \citep{Mahapatra17}. \cite{Hammond17} also investigated the possibility of clouds and found that given the observed high equilibrium temperature and a range of partial pressures from \cite{Miguel11}, Na is unlikely to condense and form clouds, whereas SiO could potentially condense on the planet's night-side (both Na and SiO could arise from a day-side magma ocean; e.g. \citealt[][]{Schaefer10,Miguel11,Hammond17}).}

\newpage

\subsection{Refraction}
{Briefly, we note that while the effects of refraction were not considered in our model atmosphere calculations (see Section \ref{subsec:models}), it has been shown that refraction can act to mute spectral features in the lower atmosphere by creating a grey continuum similar to that produced by optically thick clouds \citep{Betremieux18}. For thin atmospheres around terrestrial exoplanets, the largest pressure that can be probed may indeed be the exoplanet's surface pressure; with increasingly thicker atmospheres, however, refraction may prevent observations of the lower atmosphere.}

{\cite{Betremieux18} calculate the refractive boundaries of several hot Jupiters and terrestrial exoplanets for various atmospheric compositions in order to assess the impact of refraction on observations. In most cases, these refractive boundaries are located at pressures of $> 1$ bar.}

{In the case of 55 Cancri e, which has a high equilibrium temperature ($\sim 2400$ K, \citealt{Demory16b}), and the observations/models presented in this paper, refraction should not have a large effect on our results. However, future work (particularly of observations that probe deeper into the atmosphere) may benefit from including the effects of refraction in model calculations.}

\section{Conclusion} \label{sec:conclusion}
We have presented our analysis of high-resolution near-infrared transmission spectroscopy of the transiting super-Earth 55~Cnc~e. This paper shows the results of the Doppler cross-correlation technique, which takes advantage of the large change in radial velocity of the exoplanet during its transit as well as the high spectral resolution of our observations, allowing us to resolve individual molecular features and disentangle the planetary signal from stellar and telluric absorption lines. 

In the cases of atmospheric HCN, NH${}_3$, and C${}_2$H${}_2$, we are able to place strong upper limits. We can rule HCN out of the atmosphere of 55~Cnc~e at a volume mixing ratio as low as 0.001\% with a mean molecular weight of 2 amu; if the mean molecular weight is increased to 5 amu, we can rule HCN out down to a volume mixing ratio of 0.02\%. Our findings rule out the most likely models suggested by the analysis of \cite{Tsiaras16}, but there remain several models with lower volume mixing ratios and a mean molecular weight of $\sim$4 or 5 amu that are both consistent with the \cite{Tsiaras16} analysis and are not ruled out by our observations.

We can rule NH${}_3$ out of the atmosphere at a volume mixing ratio as low as 0.0025\% if the mean molecular weight is 2 amu; if it is increased to 5 amu, we can rule NH${}_3$ out down to a volume mixing ratio of 0.08\%. Recent work suggests that the atmosphere of 55~Cnc~e is likely N-dominated \citep{Angelo17, Hammond17}. Moreover, \cite{Miguel19} showed that both NH${}_3$ and HCN should be present in an N-dominated atmosphere, and that transmission spectra from 3 -- 20 $\mu$m should show strong features of NH${}_3$ and HCN. Our results rule out low-mean-molecular-weight, high-volume-mixing-ratio scenarios for both molecules; however, they are consistent with calculations suggesting that the atmosphere should have a high mean molecular weight, thus with the conclusions of \cite{Miguel19}.

Finally, we can rule C${}_2$H${}_2$ out of the atmosphere at a volume mixing ratio as low as 0.08\% if the mean molecular weight is 2 amu; if the mean molecular weight is increased to 5 amu, we can rule C${}_2$H${}_2$ out of the atmosphere down to a volume mixing ratio of 1.0\%.

In the case of atmospheric CO, CO${}_2$, and H${}_2$O, on the other hand, we are unable to place significant constraints. We note that while the injections of CO models with a VMR of either 10\% or 1\% and a mean molecular weight of 2 amu do result in $>3\sigma$ detections, there are many additional features in the data at $>1\sigma$, and we thus caution that the peaks seen at $>3\sigma$ in Fig. \ref{fig:co} should be treated as tentative and warranting further investigation.

Through several tests designed to probe the limits of our detection capabilities, we conclude that our inability to recover injected H${}_2$O and CO${}_2$ models stems from the difficulty of removing telluric absorption lines across the broad NIR wavelength range of our data. Future analyses would benefit from exploring additional avenues of telluric correction.

While we do not detect an atmosphere around 55~Cnc~e, we do provide improved constraints on the possible scenarios for one that may exist. Furthermore, we have demonstrated the efficacy of the Doppler cross-correlation method at detecting nitrogen-rich molecules in particular, which will be of use in future studies of super-Earth atmospheres. Our understanding of the nature of 55~Cnc~e's atmosphere is likely to advance in the coming years, thanks to the increased wavelength coverage of upcoming spectrographs that extend further into the infrared, improved telluric absorption removal techniques, and the launch of the James Webb Space Telescope.

\acknowledgements

We wish to thank Miranda Herman, Yamila Miguel, and Tom\'{a}s Cassanelli for the helpful discussions. {We also wish to thank the anonymous referee for their kind and thorough evaluation, which greatly strengthened this work. Finally, we thank the scientific editor Michael Endl and the anonymous data editor for their helpful input.}

E.K.D was supported by an NSERC Canada Graduate Scholarship-Master's and a Vanier Canada Graduate Scholarship.

We acknowledge that this work was conducted on the traditional land of the Huron-Wendat, the Seneca, and most recently, the Mississaugas of the Credit River; we are grateful for the
opportunity to work here. We also wish to acknowledge that this work is based on observations obtained at the Canada-France-Hawaii Telescope (CFHT) which is operated from the summit of Maunakea by the National Research Council of Canada, the Institut National des Sciences de l'Univers of the Centre National de la Recherche Scientifique of France, and the University of Hawaii. The observations at the Canada-France-Hawaii Telescope were performed with care and respect from the summit of Maunakea, which is a significant cultural and historic site. This work is also based on observations collected at the Centro Astron\'{o}mico Hispano Alem\'{a}n (CAHA), operated jointly by the Max-Planck Institut f\"{u}r Astronomie and the Instituto de Astrof\'{i}sica de Andalucia (CSIC).

\software{astropy \citep{astropy:2013, astropy:2018}, Numpy \citep{numpy11, harris2020},
CARACAL \citep{Caballero16},
Molecfit \citep{Smette15, Kausch15},
occultquad \citep{Mandel2002},
Matplotlib \citep{Hunter:2007},
SciPy \citep{2020SciPy-NMeth},
IPython \citep{4160251}
}

\appendix 
\section{Data Reduction}
\label{app:reduc}
The full data reduction process for each night and each order of our observations is presented here. The process is described in Section \ref{sec:reduc}, and examples of the process being applied to specific orders of the CARMENES and SPIRou datasets are shown in the left and right panels of Fig. \ref{fig:reduction} respectively. The orders containing significant telluric contamination (i.e. those between the standard photometric bands) were excluded from further analysis, and are indicated by a greyscale in the figures.

\subsection{CARMENES}
\begin{figure*}
    \centering
    \includegraphics{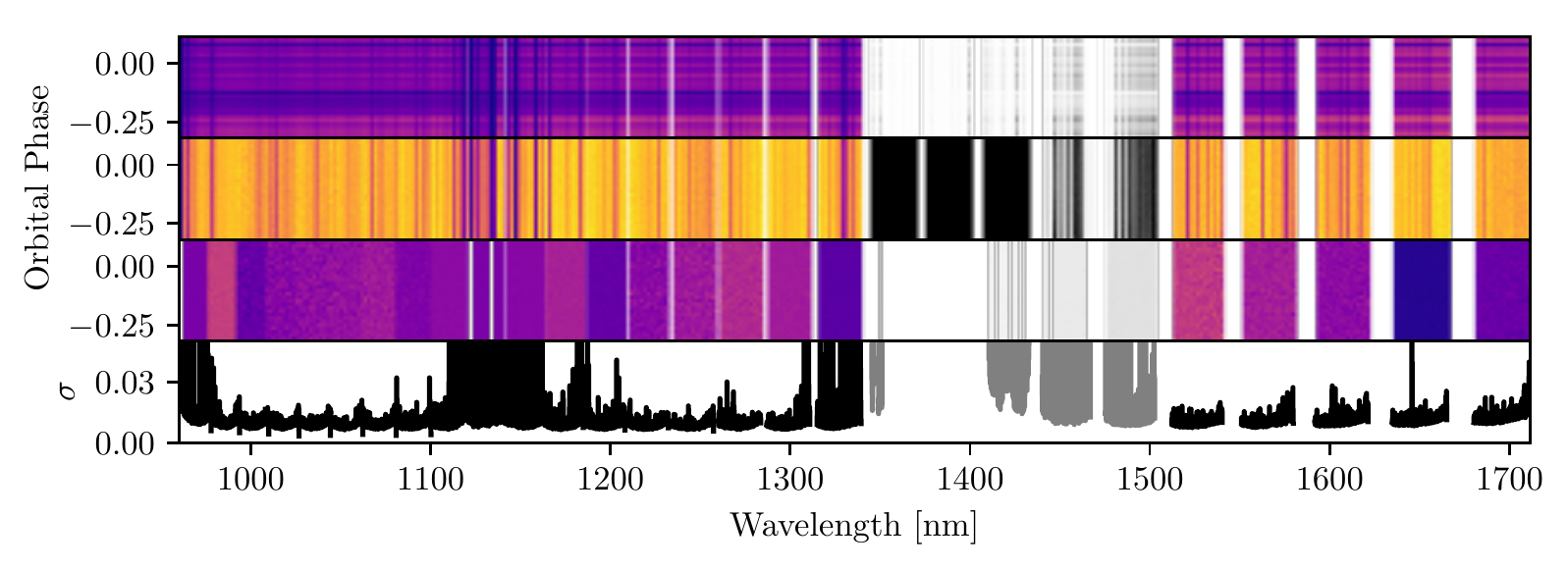}
    \caption{The data reduction process as applied to the first night of CARMENES observations (N${}_1$ in Table \ref{tab:obs}). The top panels show the data after they have been extracted from and reduced by the telescope, the second panels show the data after applying a blaze correction and median filtering algorithm (as described in Section \ref{sec:reduc}), the third panels show the data after applying 6 iterations of the \textsc{sysrem} algorithm (see Section 
    \ref{subsec:sysrem}), and the fourth panels show the standard deviation along each wavelength channel after applying the \textsc{sysrem} algorithm.}
    \label{fig:carmenes_n1}
\end{figure*}

\begin{figure*}
    \centering
    \includegraphics{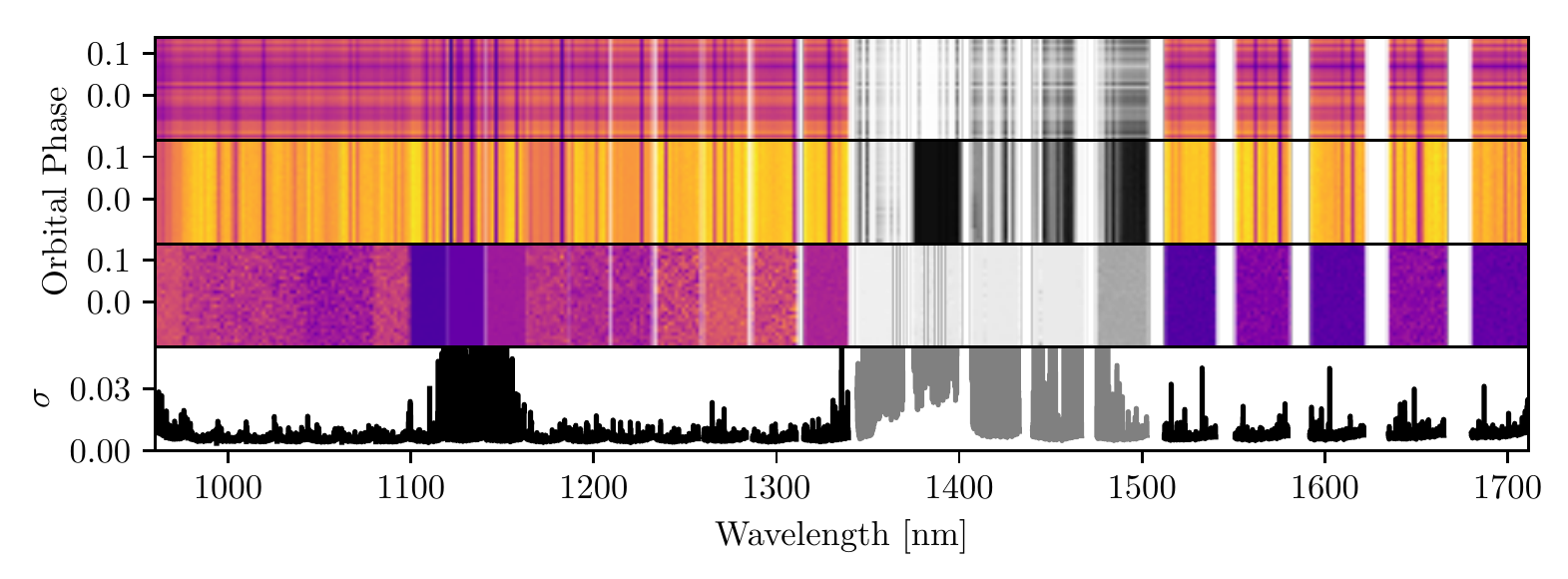}
    \caption{The data reduction process as applied to the second night of CARMENES observations (N${}_2$ in Table \ref{tab:obs}). The panels are as described in the caption of Fig. \ref{fig:carmenes_n1}.}
    \label{fig:carmenes_n2}
\end{figure*}

\begin{figure*}
    \centering
    \includegraphics{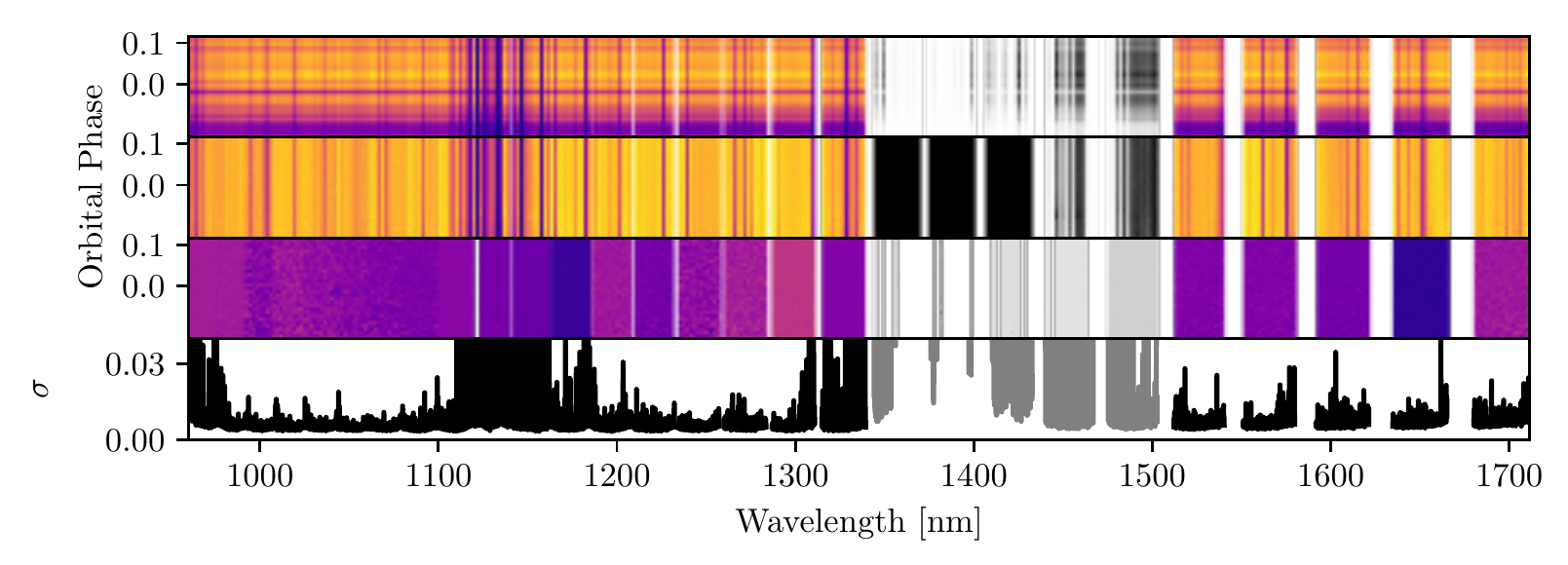}
    \caption{The data reduction process as applied to the third night of CARMENES observations (N${}_3$ in Table \ref{tab:obs}). The panels are as described in the caption of Fig. \ref{fig:carmenes_n1}.}
    \label{fig:carmenes_n3}
\end{figure*}

\begin{figure*}
    \centering
    \includegraphics{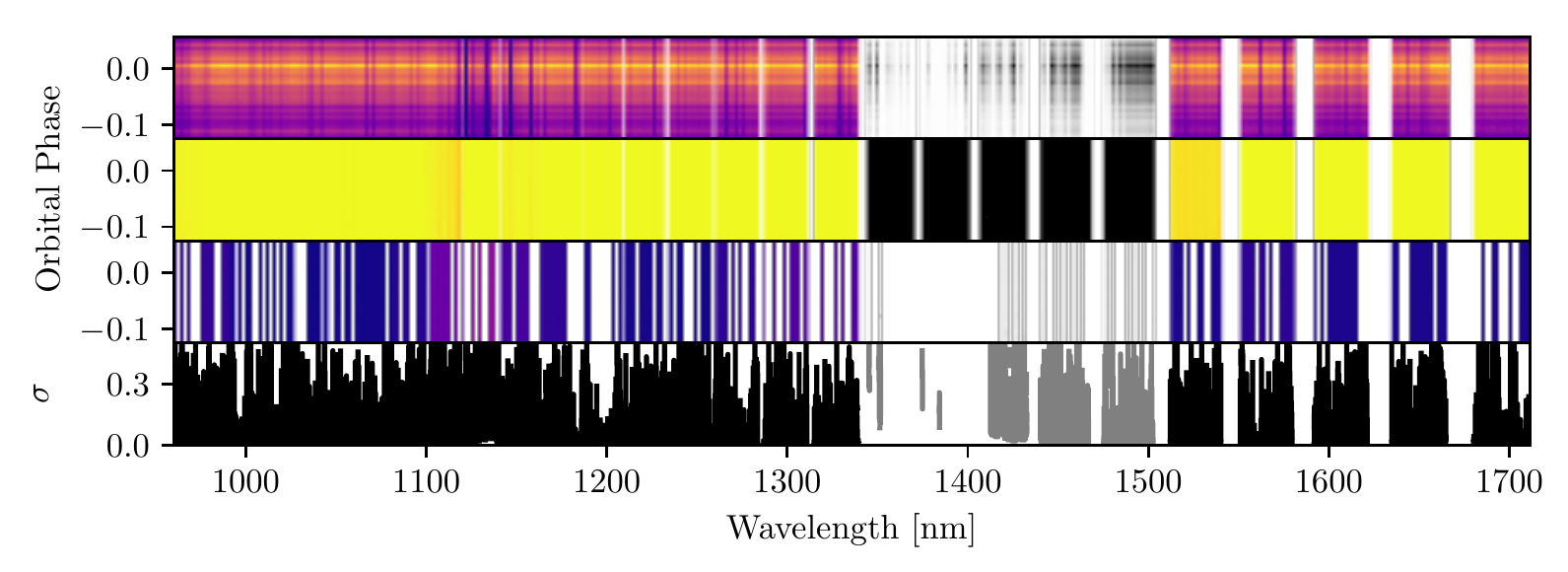}
    \caption{The data reduction process as applied to the fifth night of CARMENES observations (N${}_5$ in Table \ref{tab:obs}). The panels are as described in the caption of Fig. \ref{fig:carmenes_n1}. We note that \textsc{sysrem} performed poorly in many orders as seen in the third and fourth panels of the plot; however, we have chosen to keep these observations in our analysis. As described in Section \ref{sec:analysis}, we weight each wavelength channel by its standard deviation, and thus we ensure that the areas where \textsc{sysrem} performed poorly do not negatively impact our results.}
    \label{fig:carmenes_n5}
\end{figure*}

Figs. \ref{fig:carmenes_n1}, \ref{fig:carmenes_n2}, \ref{fig:carmenes_n3}, and \ref{fig:carmenes_n5} show the results of applying the full data reduction process to the four nights of CARMENES observations used in this analysis. We chose to exclude several orders from all further analyses due to severe telluric contamination that prevented the blaze correction or the \textsc{sysrem} algorithm from being able to provide an adequate reduction; these are seen in regions where the standard deviation is much higher than the rest of the spectrum. Note that we also chose to exclude N${}_4$ and N${}_6$ from our analysis due to poor observing conditions (see Table \ref{tab:obs}).

\subsection{SPIRou}
\begin{figure*}
    \centering
    \includegraphics{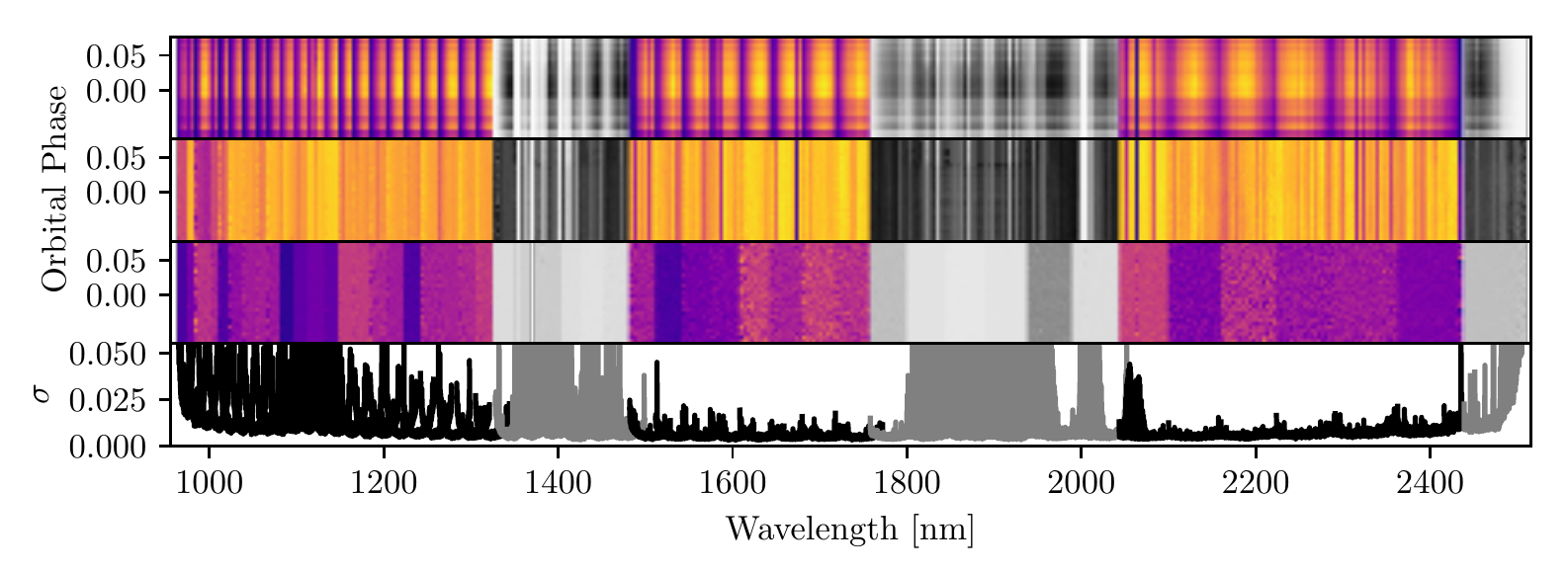}
    \caption{The data reduction process as applied to the first night of SPIRou observations (N${}_7$ in Table \ref{tab:obs}). The panels are as described in the caption of Fig. \ref{fig:carmenes_n1}.}
    \label{fig:spirou_n1}
\end{figure*}

\begin{figure*}
    \centering
    \includegraphics{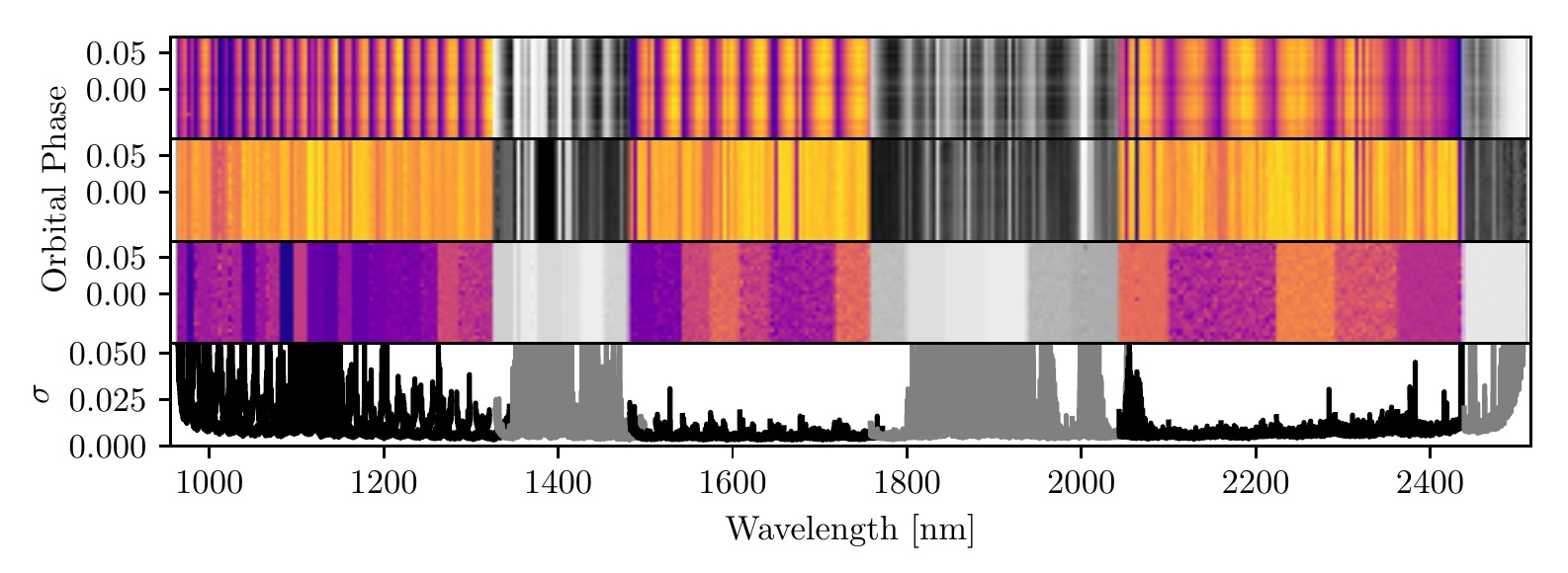}
    \caption{The data reduction process as applied to the second night of SPIRou observations (N${}_8$ in Table \ref{tab:obs}). The panels are as described in the caption of Fig. \ref{fig:carmenes_n1}.}
    \label{fig:spirou_n2}
\end{figure*}

\begin{figure*}
    \centering
    \includegraphics{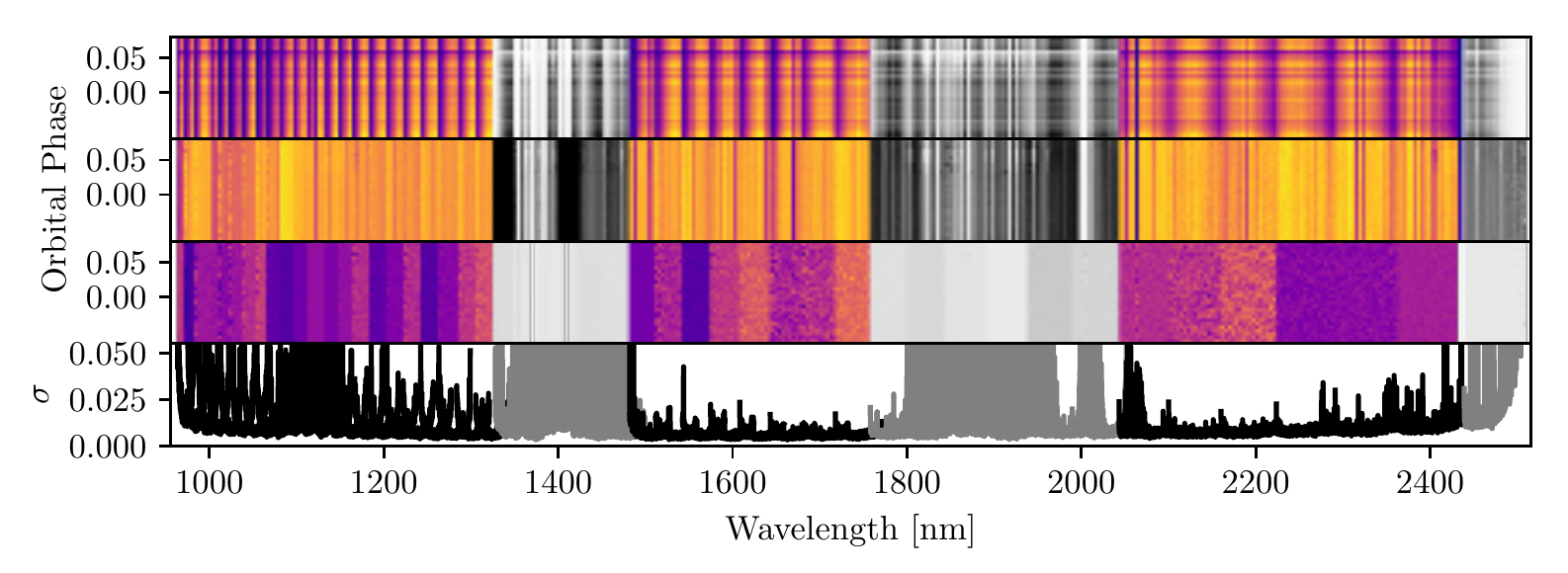}
    \caption{The data reduction process as applied to the third night of SPIRou observations (N${}_9$ in Table \ref{tab:obs}). The panels are as described in the caption of Fig. \ref{fig:carmenes_n1}.}
    \label{fig:spirou_n3}
\end{figure*}

\begin{figure*}
    \centering
    \includegraphics{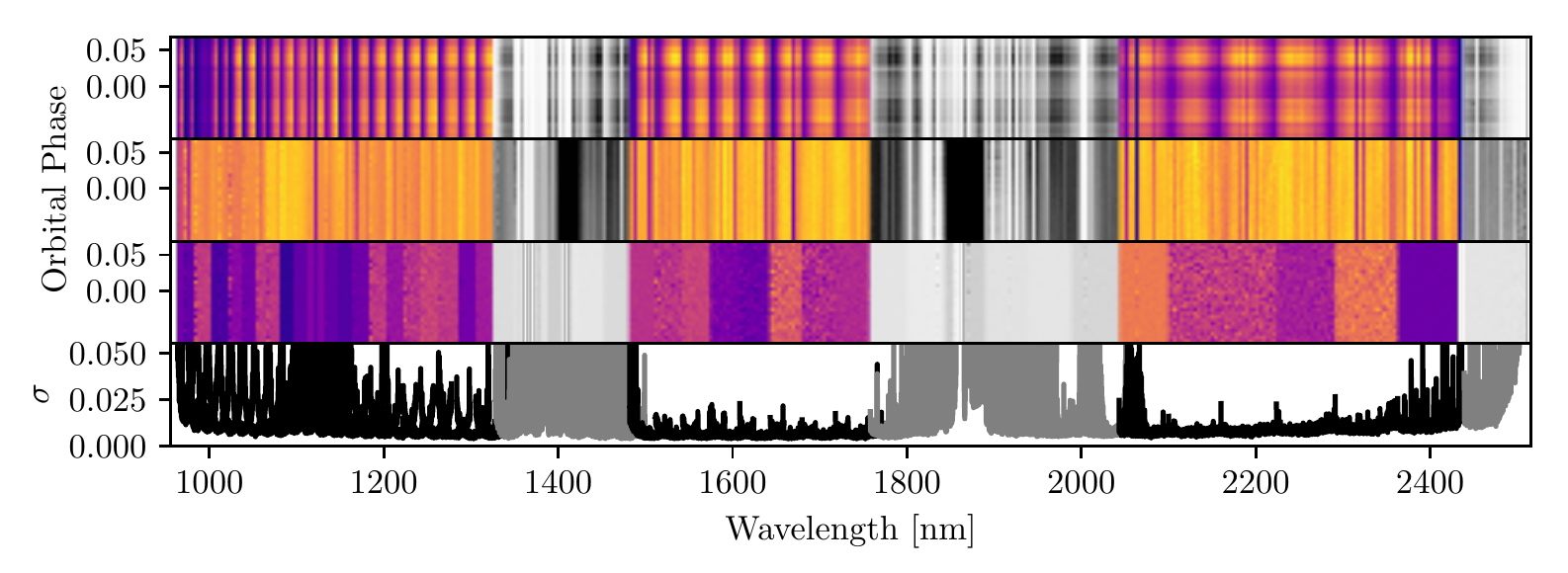}
    \caption{The data reduction process as applied to the fourth night of SPIRou observations (N${}_{10}$ in Table \ref{tab:obs}). The panels are as described in the caption of Fig. \ref{fig:carmenes_n1}.}
    \label{fig:spirou_n4}
\end{figure*}

Figs. \ref{fig:spirou_n1}, \ref{fig:spirou_n2}, \ref{fig:spirou_n3}, and \ref{fig:spirou_n4} show the results of applying the full data reduction process to the four nights of SPIRou observations. Again, we have excluded regions of severe telluric contamination from our analysis and indicated these in greyscale.

\section{Models}
\label{app:models}
\begin{figure*}
    \centering
    \includegraphics{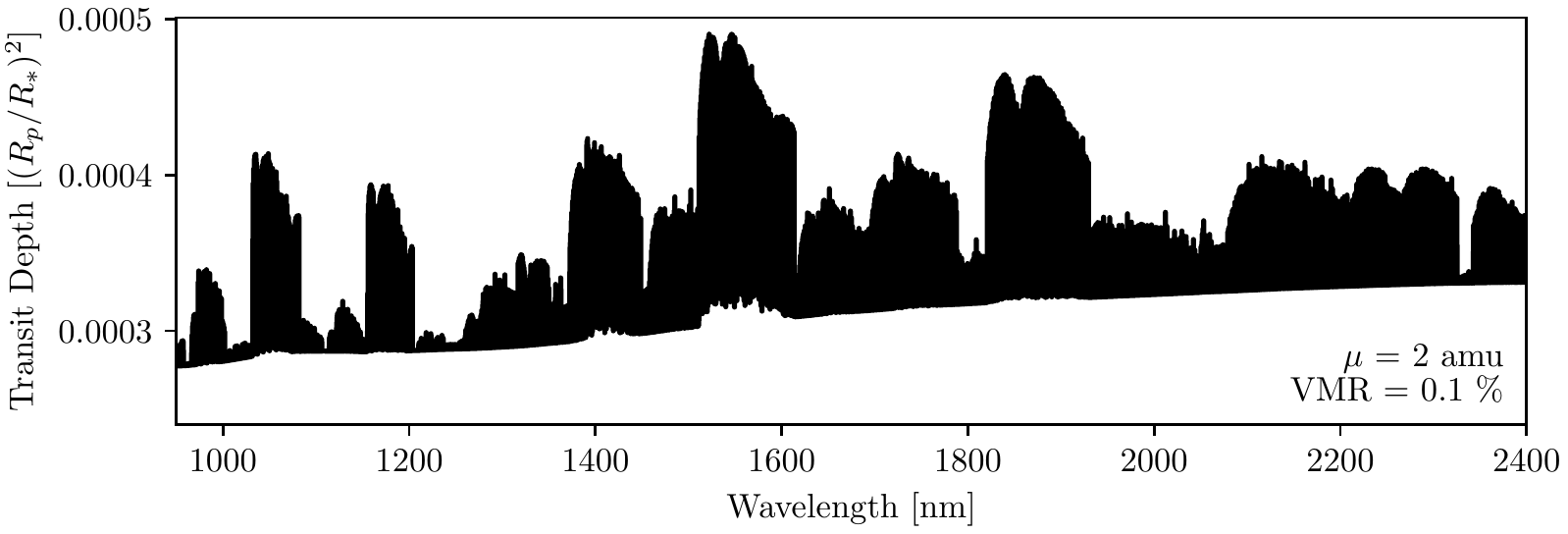}
    \caption{An example HCN model calculated for 55~Cnc~e with a mean molecular weight of 2 amu and a volume mixing ratio of 0.1\%. This model is ruled out of the atmosphere, as seen in the top-left panel of Fig. \ref{fig:hcn_result}.}
    \label{fig:hcn_examplemodel}
\end{figure*}

\begin{figure*}
    \centering
    \includegraphics{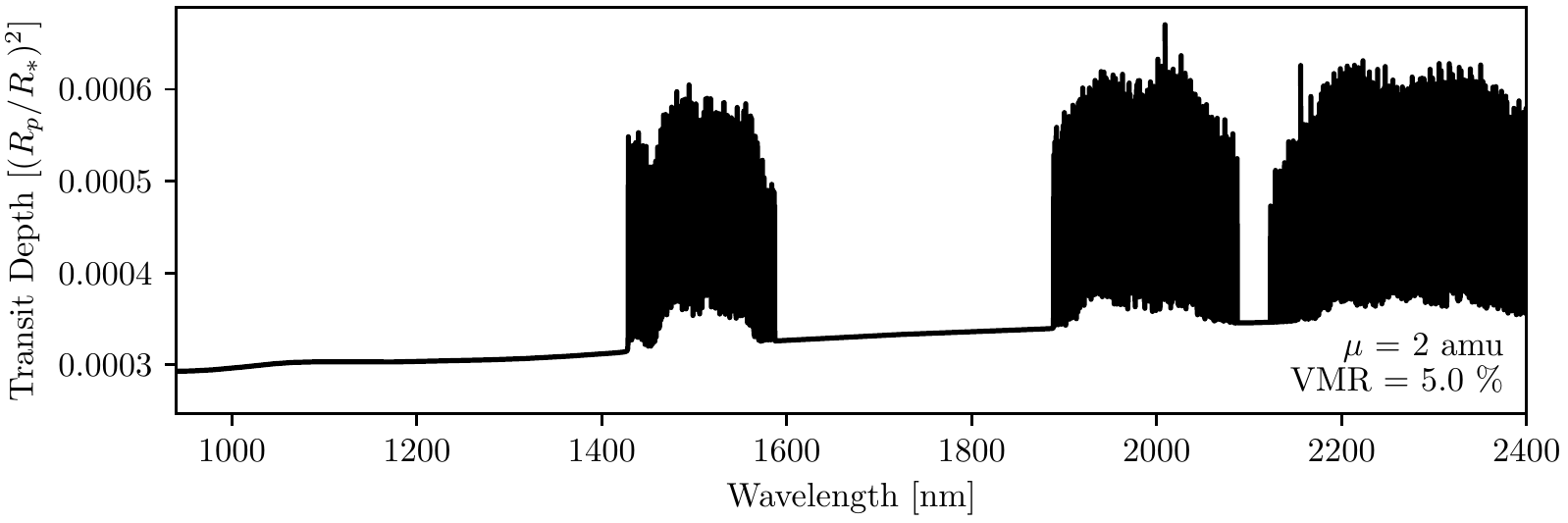}
    \caption{An example NH${}_3$ model calculated for 55~Cnc~e with a mean molecular weight of 2 amu and a volume mixing ratio of 5.0\%. This model is ruled out of the atmosphere, as seen in the top-left panel of Fig. \ref{fig:nh3_result}.}
    \label{fig:nh3_examplemodel}
\end{figure*}

\begin{figure*}
    \centering
    \includegraphics{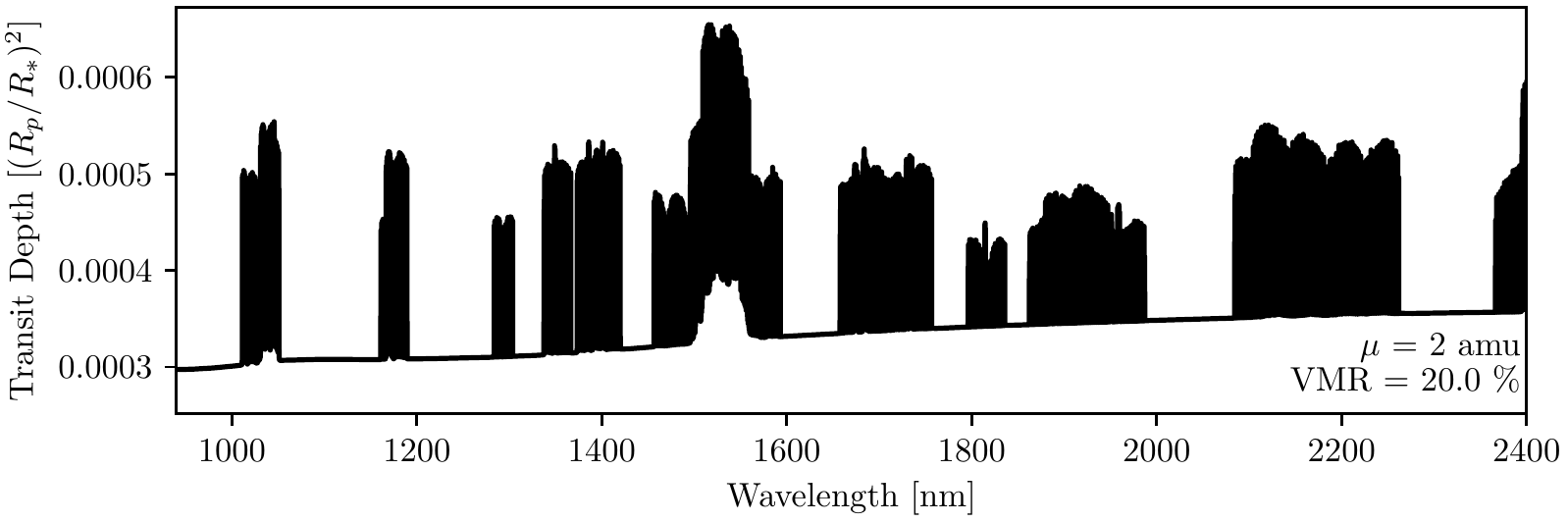}
    \caption{An example C${}_2$H${}_2$ model calculated for 55~Cnc~e with a mean molecular weight of 2 amu and a volume mixing ratio of 20\%. This model is ruled out of the atmosphere, as seen in the top-left panel of Fig. \ref{fig:c2h2_result}.}
    \label{fig:c2h2_examplemodel}
\end{figure*}

\begin{figure*}
    \centering
    \includegraphics{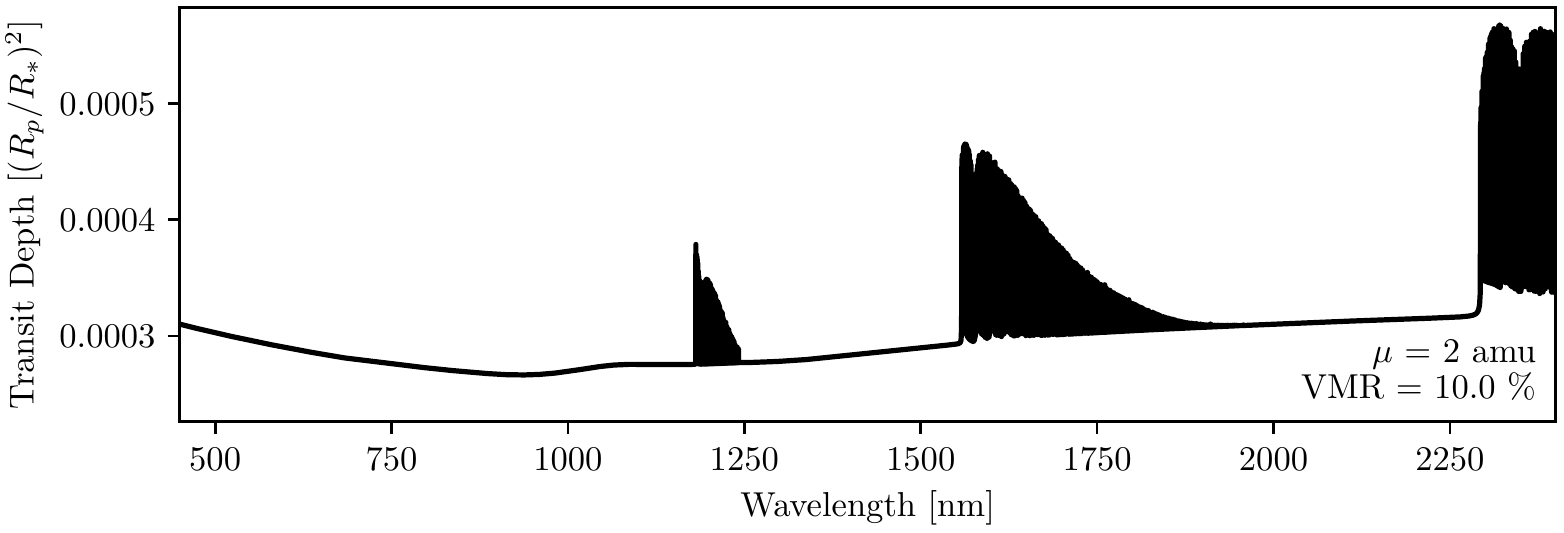}
    \caption{An example CO model calculated for 55~Cnc~e with a mean molecular weight of 2 amu and a volume mixing ratio of 10\%. This model is tentatively ruled out of the atmosphere, as seen in the top-left panel of Fig. \ref{fig:co}. We caution that this result is only tentative, as there are numerous additional peaks at $>1\sigma$ in our results.}
    \label{fig:co_examplemodel}
\end{figure*}

\begin{figure*}
    \centering
    \includegraphics{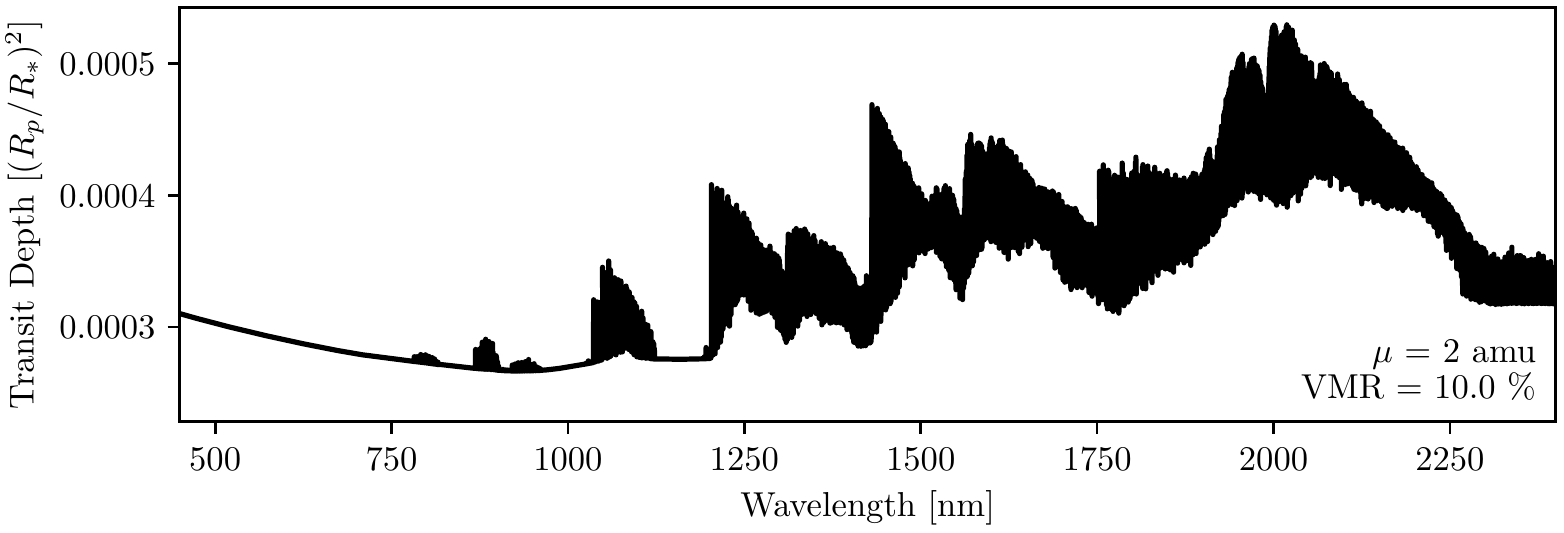}
    \caption{An example CO${}_2$ model calculated for 55~Cnc~e with a mean molecular weight of 2 amu and a volume mixing ratio of 10\%. This model corresponds to the phase-folded cross correlations shown in the top-left panel of Fig. \ref{fig:co2_result}, and is not ruled out of the planet's atmosphere by our analysis.}
    \label{fig:co2_examplemodel}
\end{figure*}

\begin{figure*}
    \centering
    \includegraphics{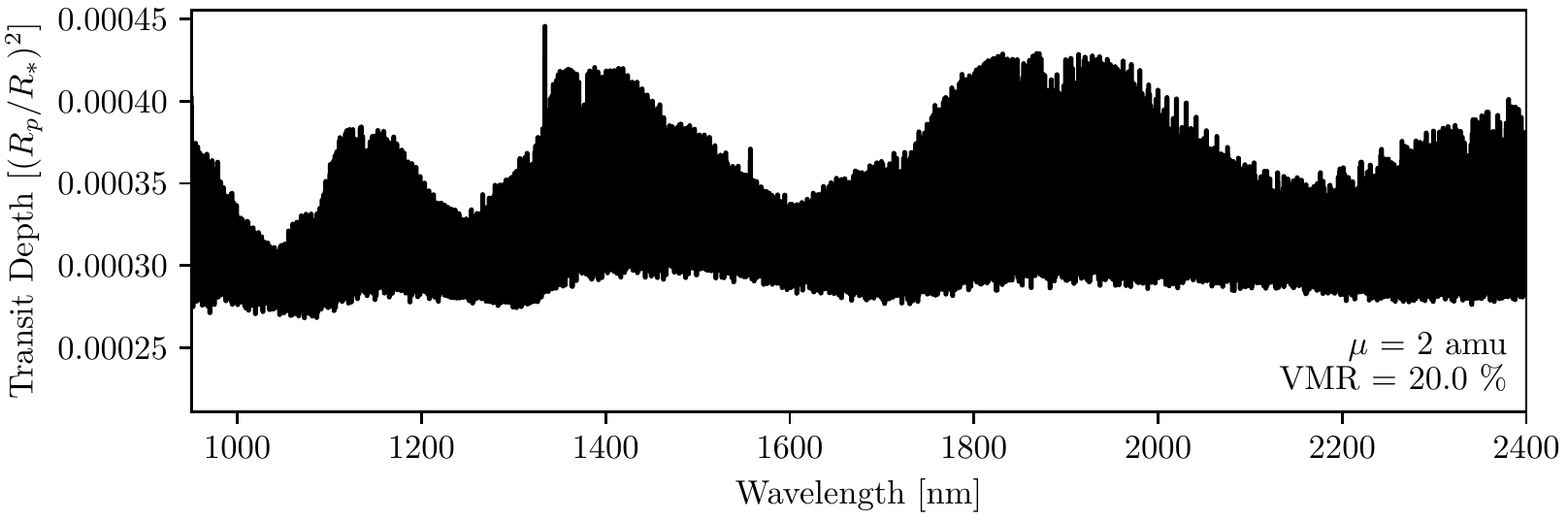}
    \caption{An example H${}_2$O model calculated for 55~Cnc~e with a mean molecular weight of 2 amu and a volume mixing ratio of 20\%. This model corresponds to the phase-folded correlations shown in the top-left panel of Fig. \ref{fig:h2o_result}, and is not ruled out of the planet's atmosphere by our analysis.}
    \label{fig:h2o_examplemodel}
\end{figure*}

In this section we present example models for each molecule considered in our analysis (HCN, NH${}_3$, C${}_2$H${}_2$, CO, CO${}_2$, and H${}_2$O; Figs. \ref{fig:hcn_examplemodel}, \ref{fig:nh3_examplemodel}, \ref{fig:c2h2_examplemodel}, \ref{fig:co_examplemodel}, \ref{fig:co2_examplemodel}, and \ref{fig:h2o_examplemodel}, respectively). Further details on these models, including a description of how they were generated, are available in Section \ref{subsec:models}. Each example model shown here corresponds to the top-left model of each plot in Section \ref{sec:discussion}; e.g. Figs. \ref{fig:hcn_result}, \ref{fig:nh3_result}, \ref{fig:c2h2_result}, \ref{fig:co}, \ref{fig:co2_result}, and \ref{fig:h2o_result}.

{For some of the models presented in this work (e.g. the C${}_2$H${}_2$ model in Fig. \ref{fig:c2h2_examplemodel}), the line lists that are currently available are incomplete. This results in the sharp cutoffs in wavelength visible in numerous bands. Going forward, we expect that more complete and accurate line list data will become available, which would allow for a more complete and realistic model to be calculated.}

\section{Molecfit}
\label{app:molecfit}
In this section, we present additional details on our telluric feature removal process using Molecfit. Although our analysis removes telluric features through the use of the \textsc{sysrem} algorithm, we decided to compare our results with Molecfit in order to determine whether the removal of telluric features was limiting our ability to recover certain injected models (see Section \ref{subsec:h2o} for further details). We largely followed the methods laid out in \cite{Allart17} and \cite{Salz18} to apply Molecfit to our observations.

\subsection{Fit Parameters}
\label{subapp:params}
The initial parameters used with Molecfit for every night are shown in Table \ref{tab:molecfit}. Further details on these parameters are available in \cite{Smette15,Kausch15}. We allowed Molecfit to fit for a Gaussian kernel that varies with wavelength, and included fitting for the wavelength solution.

\begin{deluxetable}{lcc}[htbp!]
\tablecaption{Initial parameters used by Molecfit. {We allow Molecfit to fit for molecules with absorption features present in the wavelength range of our data (see Molecules row below), and apply a second-degree Chebyschev polynomial fit to the wavelength solution ($n_\lambda$ below). The continuum (cont${}_n$ below) is fit with a second-degree polynomial, and as the data are normalized, we set an initial constant term of 1 for the continuum which is then refined through the fit. The instrumental profile is initially assumed to be a Gaussian with a FWHM of 3.5 pixels; however, this too is refined through the fit. Molecfit uses the Levenberg-Marquardt technique to quickly solve the least-squares problem \citep{Smette15}; the $\chi^2$ convergence parameter (ftol below) as well as the parameter convergence criterion (xtol below) of the Levenberg-Marquardt technique are set to $10^{-10}$.}}
\tablehead{%
    \colhead{Initial parameters} & \colhead{Value} & \colhead{Notes}
    }
\startdata
ftol & $10^{-10}$ & Relative $\chi^2$ convergence criterion \\
xtol & $10^{-10}$ & Relative parameter convergence criterion \\
Molecules & H${}_2$O, CO${}_2$, O${}_3$, CO CH${}_4$, O${}_2$ & Molecules included in the fit \\
cont${}_n$ & 2 & Degree of coefficients for continuum fit \\
cont${}_\mathrm{const.}$ & 1 & Initial constant term for continuum fit \\
$n_\lambda$ & 2 & Polynomial degree of the refined wavelength solution \\
$\omega_\mathrm{Gaussian}$ & 3.5 & Initial value for FWHM of Gaussian in pixels \\
Kernel Size & 15 & Size of Gaussian kernel in FWHM
\label{tab:molecfit}
\enddata
\end{deluxetable}

\section{Assessing our H${}_2$O and CO${}_2$ Retrieval Capabilities}
\label{app:tests}
In the following sections, we present several additional tests designed to evaluate our H${}_2$O and CO${}_2$ model recovery capabilities.

\subsection{White Noise Test}
\label{subsubsec:whitenoise}
\begin{figure*}
    \centering
    \includegraphics{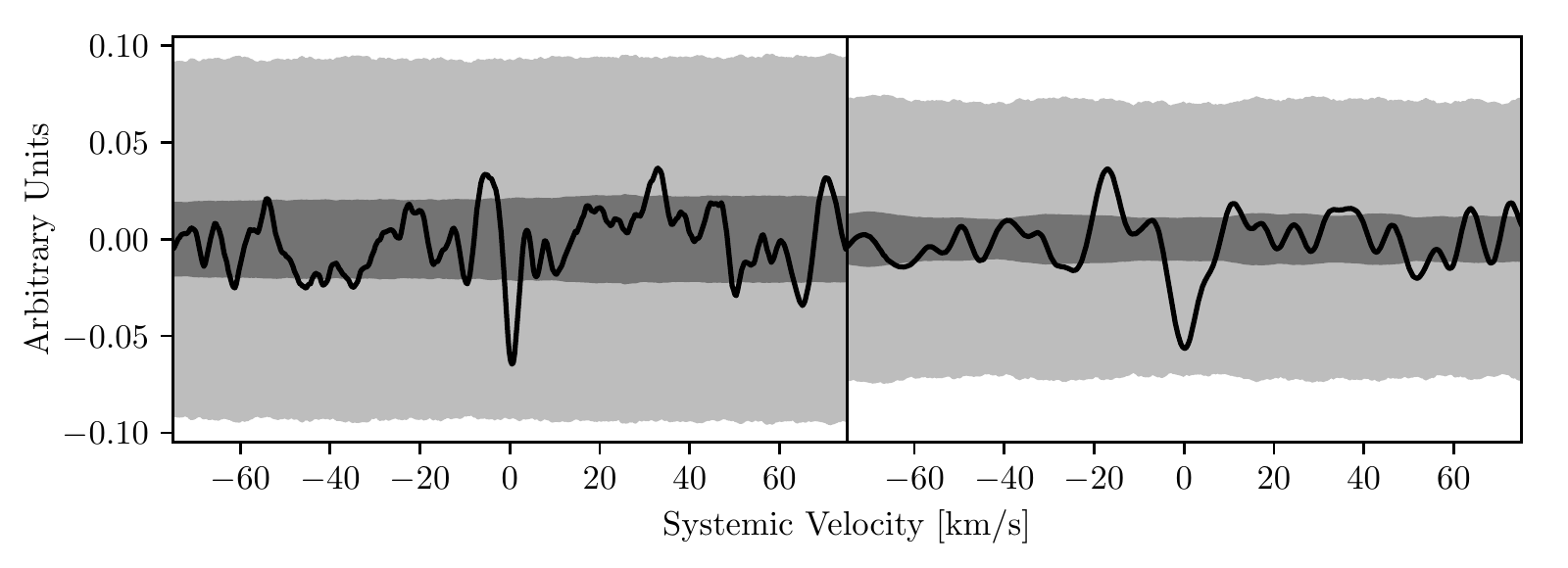}
    \caption{The white noise test for the CARMENES data, as described in Appendix \ref{subsubsec:whitenoise}. The panel on the left shows the results of injecting an atmospheric H${}_2$O model with a VMR of 20\% and a mean molecular weight of 2 amu (i.e. the model shown in Fig. \ref{fig:h2o_examplemodel}) into our CARMENES data set and repeating the Doppler cross-correlation process. The data with the model injected is represented by the black line, while the dark- and light-grey contours represent 1 and 3$\sigma$ confidence levels, respectively (see Section \ref{subsec:significance}) The panel on the right shows the results of injecting this same atmospheric model into a white noise dataset as described in Appendix \ref{subsubsec:whitenoise}. The model that has been injected into the white noise data set is closer to the 3$\sigma$ confidence level than that injected into the data; however, in neither case were we able to confidently retrieve the injected model.} 
    \label{fig:carmenes_whiteNoise}
\end{figure*}

\begin{figure*}
    \centering
    \includegraphics{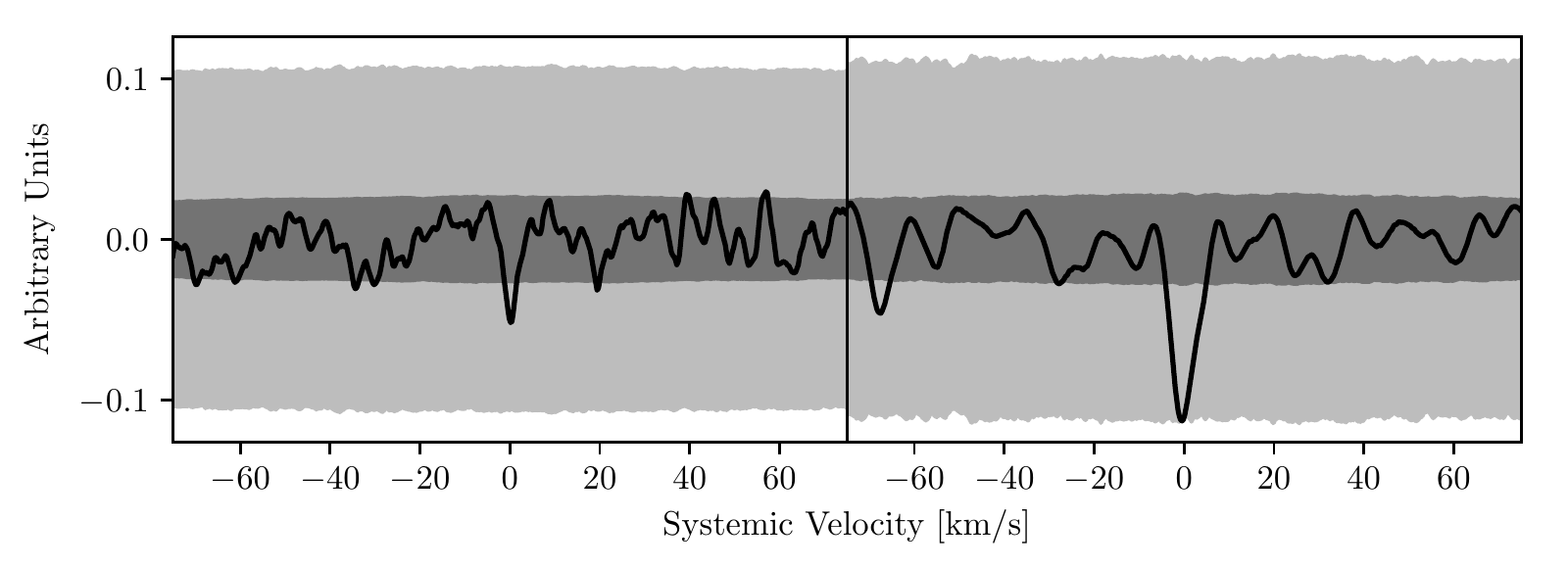}
    \caption{The same as Fig. \ref{fig:carmenes_whiteNoise}, but for a CO${}_2$ model with a VMR of 10\% and a mean molecular weight of 2 amu (i.e. the model shown in Fig. \ref{fig:co2_examplemodel}). We were unable to confidently recover the injected model in the original data set (left) or the white noise data set (right), but we note that the injected model is closer to the 3$\sigma$ confidence level in the white noise data set than the original data set.}
    \label{fig:carmenes_whiteNoise_CO2}
\end{figure*}

\begin{figure*}
    \centering
    \includegraphics{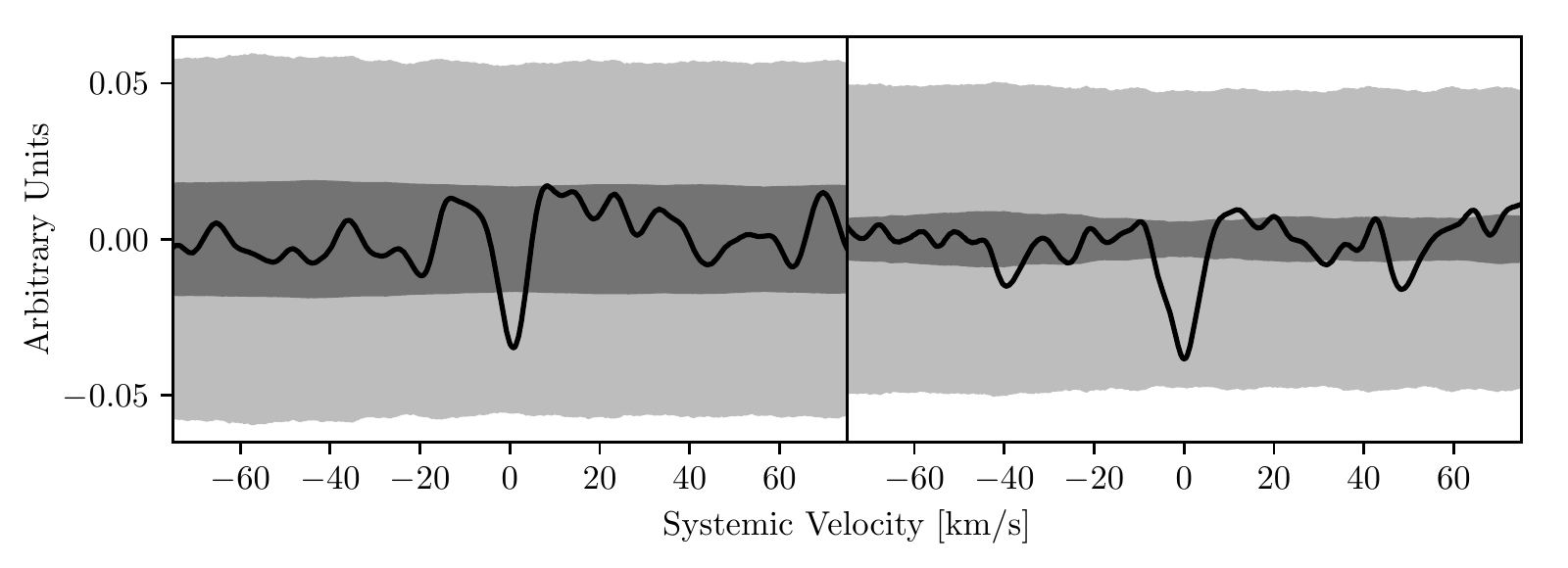}
    \caption{The white noise test for the SPIRou data, as described in Appendix \ref{subsubsec:whitenoise}. The panels are as described in the caption of Fig. \ref{fig:carmenes_whiteNoise}. Again, the model that has been injected into the white noise data set here does not pass the 3$\sigma$ confidence level.} 
    \label{fig:spirou_whiteNoise}
\end{figure*}

\begin{figure*}
    \centering
    \includegraphics{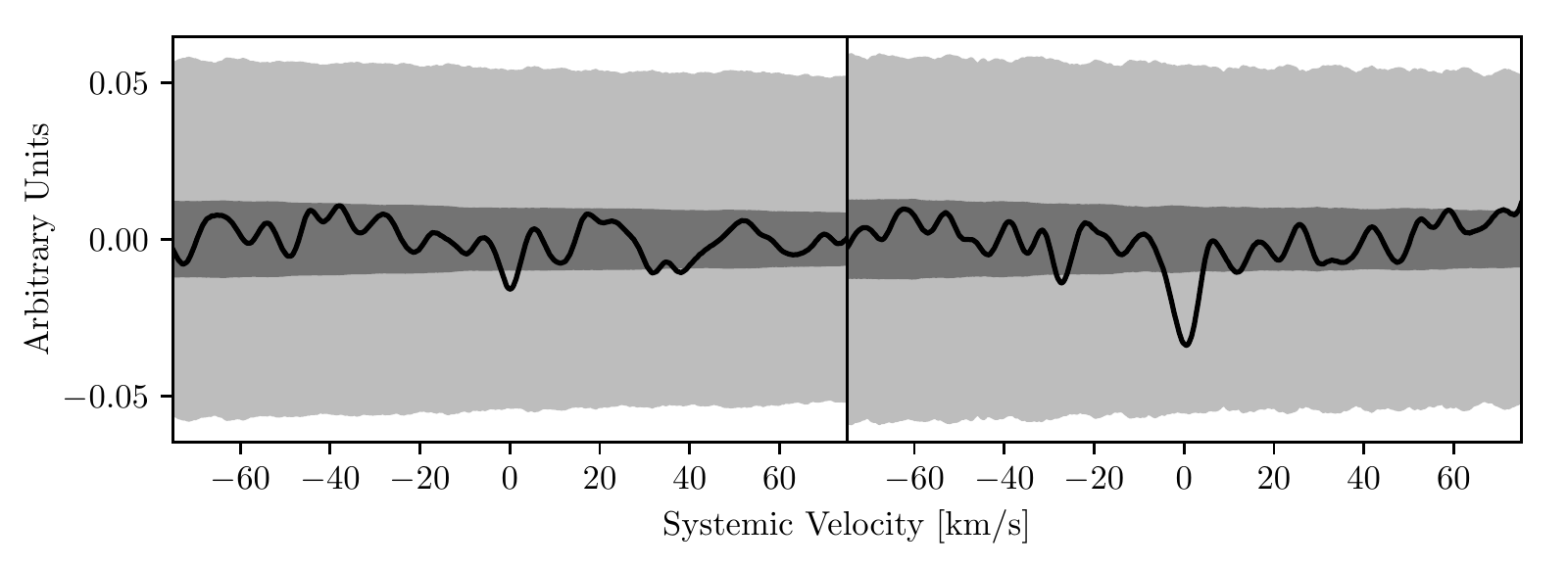}
    \caption{The same as Fig. \ref{fig:spirou_whiteNoise}, but for a CO${}_2$ model with a VMR of 10\% and a mean molecular weight of 2 amu (i.e. the model shown in Fig. \ref{fig:co2_examplemodel}). We were unable to confidently recover the injected model in the original data set (left) or the white noise data set (right). In the white noise data set, however, the injected model is closer to the 3$\sigma$ confidence level.}
    \label{fig:spirou_whiteNoise_CO2}
\end{figure*}

To test whether residual noise left in the data after the application of the \textsc{sysrem} algorithm was affecting our final result, we followed a similar method to \cite{Esteves17} by simulating a pure white noise data set generated to match the rms of each wavelength channel after applying \textsc{sysrem} (see e.g. the bottom panels of the figures in Appendix \ref{app:reduc}). We tried injecting both a model H${}_2$O atmosphere and a model CO${}_2$ atmosphere into this white noise data set and repeated our analysis routine as described in Section \ref{sec:analysis} for each model. This allowed us to assess whether the residual noise level in the data after applying the \textsc{sysrem} algorithm was the limiting factor in being able to recover H${}_2$O and CO${}_2$ models. The results of these white noise tests are shown in Figs. \ref{fig:carmenes_whiteNoise} (H${}_2$O) and \ref{fig:carmenes_whiteNoise_CO2} (CO${}_2$) for the CARMENES data set, and Figs. \ref{fig:spirou_whiteNoise} (H${}_2$O) and \ref{fig:spirou_whiteNoise_CO2} (CO${}_2$) for the SPIRou data set.

In all cases we see that in a data set comprised of pure white noise, we are unable to recover even the strongest H${}_2$O and CO${}_2$ models. This suggests that \textsc{sysrem} did not adequately remove stellar and telluric contamination in the data, and that the data themselves are simply too noisy to allow us to place strong constraints on the presence of H${}_2$O or CO${}_2$ in the atmosphere of 55~Cnc~e. We note that H${}_2$O and CO${}_2$ contain a significant number of lines throughout the broad wavelength coverages offered by CARMENES and SPIRou; while these lines can in theory improve the strength of our detections, there will also be corresponding H${}_2$O and CO${}_2$ lines present in this wavelength range in the Earth's atmosphere that contaminate the data and pose a challenge to the \textsc{sysrem} algorithm. Future analyses would benefit from exploring other avenues of telluric absorption line removal. While we have presented a brief investigation of the efficacy of Molecfit on these data (see Sections \ref{subsubsec:molec} and \ref{subsec:comparison}, as well as Appendix \ref{app:molecfit}), we note that additional insight into the strengths of various telluric removal methods on SPIRou observations specifically would be useful. In the case of CRIRES data, we note that \cite{UlmerMoll19} showed synthetic transmission methods to be preferable to a standard star method.

\subsection{Wavelength Shift}
In addition to the white noise test described in Section \ref{subsubsec:whitenoise}, we investigated whether or not a shift was present in the wavelength solution of the data. Such a shift could potentially impact our ability to recover an injected model. To test for this, we cross-correlated each spectrum of each night with the first spectrum of the night, and used a centroid algorithm to fit for the peak of the cross-correlation function. We then compared the difference in measured peaks for each cross-correlation. This was done on a per-order basis.

In the case of the CARMENES data, we find that for each night the majority of orders do not experience a large shift in the wavelength solution. For most orders, the shift in the fitted centroid of the cross-correlation function is on average 0.04 pixels. For each night we found that two orders experienced a slightly larger shift (on the order of $\sim$1 pixel); however, these were orders that had previously been discarded from the data due to excessive telluric contamination. The offending orders occur in the wavelength range between 1344.14 and 1400.24 nm; as can be seen in the figures presented in Appendix \ref{app:reduc}, this region of the data is heavily contaminated by tellurics.

In the case of the SPIRou data, the shift in the fitted centroid of the cross-correlation function for each order is on average 0.03 pixels. We do not find that any orders experience a significantly larger shift. 

In addition to the test described above, we also selected several individual telluric lines in each data set and fit these with a Gaussian, determining the centroid of each line by the Gaussian's expected value $\mu$. For these individual lines, we also did not find a significant shift in the wavelength solution.

Finally, we note that our data reduction procedure (described in Section \ref{sec:reduc}) involves interpolating each frame to a common wavelength grid.

Overall, we do not think that a shift in the wavelength solution impacted our final results, given that the shift measured by the methods described above is significantly less than 1 pixel, and less than the widths of the absorption lines themselves.

\subsection{Varying \textsc{sysrem} Iterations}
\label{subsubsec:sysremtest}
In addition to the tests described in the preceding sections, we investigated whether applying different numbers of iterations of \textsc{sysrem} affected our ability to recover injected models. While the methods we used to determine the optimal number of iterations were model-free (see Section \ref{subsec:sysrem}), it is possible that \textsc{sysrem} had begun to overfit the data and remove signal from the injected model (see for e.g. the discussion in \citealt{Brogi14}). For the purposes of this test we again used the water model with a VMR of 20\% and a mean molecular weight of 2 amu (i.e. the model shown in Fig. \ref{fig:h2o_examplemodel}) and the CO${}_2$ model with a VMR of 10\% and a mean molecular weight of 2 amu (i.e. the model shown in Fig. \ref{fig:co2_examplemodel}), because these models should be the easiest to detect. 

\begin{figure*}
    \centering
    \includegraphics{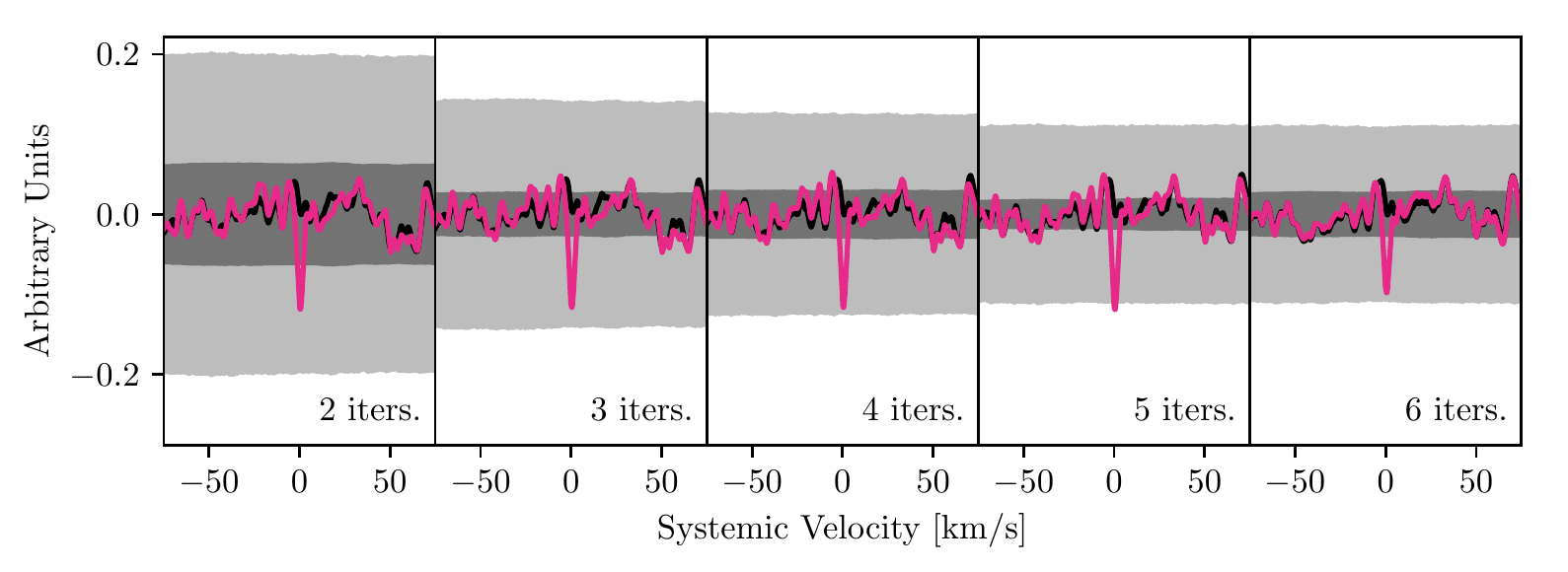}
    \caption{The results of varying the number of iterations of \textsc{sysrem} used on a data set with a water model injected, as described in Appendix \ref{subsubsec:sysremtest}. The model used had a VMR of 20\% and a mean molecular weight of 2 amu. Each panel shows a different number of iterations of the algorithm, with the number of iterations indicated in the bottom right. As in e.g. Fig. \ref{fig:h2o_result}, the black line represents the data, the magenta line represents the data with a model injected, and the dark- and light-grey contours represent 1 and 3$\sigma$ confidence levels respectively. After 5 iterations, the strength of the injected model decreases slightly.}
    \label{fig:sysrem}
\end{figure*}

\begin{figure*}
    \centering
    \includegraphics{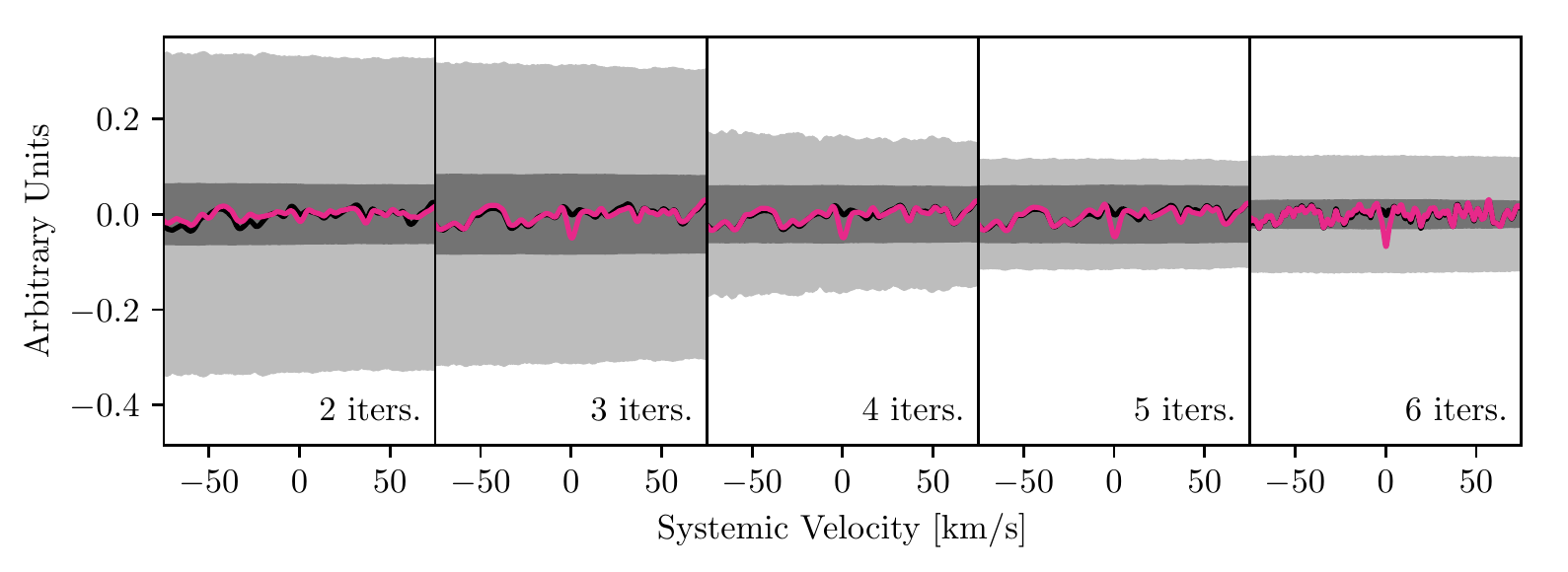}
    \caption{The results of varying the number of iterations of \textsc{sysrem} used on a data set with a carbon dioxide model injected, as described in Appendix \ref{subsubsec:sysremtest}. The model used had a VMR of 10\% and a mean molecular weight of 2 amu. The panels are as described in the caption of Fig. \ref{fig:sysrem}. After 6 iterations of \textsc{sysrem}, the model is more readily detected.}
    \label{fig:sysrem_co2}
\end{figure*}

The results are shown in Figs. \ref{fig:sysrem} and \ref{fig:sysrem_co2}. The noise level decreases with increasing numbers of iterations of \textsc{sysrem}, and appears to plateau at 6 iterations, as determined previously (see Section \ref{subsec:sysrem}).

However, Fig. \ref{fig:sysrem} also shows that the strength of the injected water model decreases slightly after 5 iterations of the algorithm: while the injected model after 5 iterations surpasses the 3$\sigma$ level, this is not the case for the injected model after 6 iterations. This indicates that the \textsc{sysrem} algorithm has begun to remove signal contributed by the model itself. In the case of CO${}_2$, however, we find that the model is more readily detected after 6 iterations.

Although this test demonstrates that varying the number of iterations of \textsc{sysrem} can have an impact on our ability to recover an injected model, we have chosen not to change the number of iterations used in our analysis for several reason. First, the number of iterations initially chosen was based on an average optimal result across all orders, and was independent of the model or molecule being considered. We have chosen to implement a model-free method for determing the optimal number of iterations of \textsc{sysrem} so that the specifics of the models being injected are less likely to impact our results.

We also note that the change in strength between 5 and 6 iterations as seen in Fig. \ref{fig:sysrem} is not highly significant: while the injected model after 5 iterations slightly surpasses the 3$\sigma$ level, the injected model after 6 iterations is close to (but not quite at) the 3$\sigma$ level. While it is possible that the number of iterations of \textsc{sysrem} used in our analysis affected some of the results shown in Fig. \ref{fig:h2o_result}, these effects are small compared to the difference in the strengths of injected models in our study and those of \cite{Jindal20}, and we thus do not consider this to be the dominating factor in our ability to recover an injected model. Instead, our ability to recover an injected model appears to be limited by the overall noise level of the data, as demonstrated by the white noise test described in Appendix \ref{subsubsec:whitenoise}. While \textsc{sysrem} doesn't appear to have adequately removed contributions from the Earth's atmosphere, it has also begun to remove the signal contributed by some injected models. Again, we believe that future analyses of data in the wavelength ranges covered by CARMENES and SPIRou would benefit from exploring alternate methods of removing stellar and telluric lines, including a detailed comparison between the capabilities of \textsc{sysrem} and Molecfit.

\bibliography{references.bib}

\end{document}